%% file: main.tex
\documentclass{article} 
\usepackage{blindtext}

\input{math_commands.tex}

\usepackage{dmlr2e}

\usepackage{lastpage}
\dmlrheading{23}{2025}{1-\pageref{LastPage}}{01/25; Revised 06/25}{07/25}{21-0000}{Schultze, Kerkfeld, Kuebart, Weber, Wolter, and Selgert} 
\def\openreview{\url{https://openreview.net/forum?id=YHDfqvtye9}} 

\ShortHeadings{Chronicling Germany}{Schultze, Kerkfeld, Kuebart, Weber, Wolter and Selgert}
\firstpageno{1}

\usepackage[utf8]{inputenc} 
\usepackage[TS1, T1]{fontenc}    
\usepackage{booktabs}       
\usepackage{nicefrac}       
\usepackage{microtype}      
\usepackage{xcolor}         
\usepackage{multirow}

\usepackage{lmodern}

\usepackage{graphicx}
\usepackage[mode=buildnew]{standalone}
\usepackage{tikz}
\usetikzlibrary{patterns,shapes.arrows,positioning,arrows,calc,matrix, fit}
\usetikzlibrary{external}
\usepackage{pgfplots}
\pgfplotsset{compat=newest}
\usepgfplotslibrary{groupplots,dateplot}
\usepackage{float}

\pgfplotsset{compat=1.18}

\definecolor{tabBlue}{HTML}{1f77b4}
\definecolor{tabOrange}{HTML}{ff7f0e}
\definecolor{tabGreen}{HTML}{2ca02c}
\definecolor{tabRed}{HTML}{d62728}
\definecolor{tabPurple}{HTML}{9467bd}
\definecolor{tabBrown}{HTML}{8c564b}
\definecolor{tabPink}{HTML}{e377c2}
\definecolor{tabGray}{HTML}{7f7f7f}
\definecolor{tabOlive}{HTML}{bcbd22}
\definecolor{tabCyan}{HTML}{17becf}

\usepackage{acro}
\acsetup{make-links=true} 
\input{acronyms.tex}

\begin{document}

    \title{Chronicling Germany: An Annotated Historical Newspaper Dataset}

    %

    \author{\name Christian Schultze \\
        \addr High-Performance Computing and Analytics (HPCA-Lab)
        Universität Bonn
        \email s6cnscul@uni-bonn.de
        \AND
        \name Niklas Kerkfeld \\
        \addr HPCA-Lab, Universität Bonn
        \email s6nikerk@uni-bonn.de
        \AND
        \name Kara Kuebart \\
        \addr Institut für Geschichtswissenschaft Universität Bonn
        \email kara.kuebart@uni-bonn.de
        \AND
        \name Princilia Weber \\
        \addr Institut für Geschichtswissenschaft, Universität Bonn
        \email s5prwebe@uni-bonn.de
        \AND
        \name Moritz Wolter$^{*}$\\
        \addr HPCA-Lab, Universität Bonn
        \email moritz.wolter@uni-bonn.de
        \AND
        \name Felix Selgert$^{*}$ \\
        \addr Institut für Geschichtswissenschaft, Universität Bonn
        \email fselgert@uni-bonn.de} 

    \editor{Hugo Jair Escalante}

    \maketitle 

    \def\thefootnote{*}\footnotetext{Equal supervision}
    \def\thefootnote{\arabic{footnote}}

    \begin{abstract}
        The correct detection of dense article layout and the recognition of characters in historical newspaper pages remains a challenging requirement for Natural Language Processing (NLP) and machine learning applications in the field of digital history. Digital newspaper portals for historic Germany typically provide Optical Character Recognition (OCR) text, albeit of varying quality. Unfortunately, layout information is often missing, limiting this rich source’s scope. Our dataset is designed to enable the training of layout and OCR models for historic German-language newspapers. The Chronicling Germany dataset contains 801 annotated historical newspaper pages from the time period between 1617 and 1933. The paper presents a processing pipeline and establishes baseline results on in- and out-of-domain test data using this pipeline. Both our dataset and the corresponding baseline code are freely available online. This work creates a starting point for future research in the field of digital history and historic German language newspaper processing. Furthermore, it provides the opportunity to study a low-resource task in computer vision.
    \end{abstract}

    \begin{keywords}
        Digital History, Machine Learning, \acf{ocr}
    \end{keywords}

    \section{Introduction}
    Newspapers are essential sources of information, not just for modern readers, but particularly in the past when
    other communication channels like the internet or radio were not yet available. Even more importantly, historical
    newspapers allow historians and social scientists to study social groups' opinions and cultural values and to use
    information from newspapers in causal models (for more details, see \ref{sec:hist_motivation}).
    This paper presents the \textit{Chronicling Germany}
    -dataset, consisting of 801 annotated high-resolution scanned newspaper pages from the period between 1617 and 1933.

    With the emergence of digital newspaper portals, using historical newspapers has become easier in recent years
    \footnote{For Germany, e.g.,
        the \textit{Deutsche Zeitungsportal} ( \url{https://www.deutsche-digitale-bibliothek.de/newspaper/}) and
        \textit{zeit.punkt NRW} (\url{https://zeitpunkt.nrw/})}.
    These portals provide text via \ac{ocr}
    but often lack reliable layout information for their German language content, which is essential for digital history applications, many of which would require newspaper articles to be treated as individual documents. Our dataset will help to reduce the character error rate and aims at considerably improving the detection of individual elements of a newspaper page, like articles or single advertisements. The former is important to prevent algorithms from connecting unrelated text regions and preserve the order in which text regions should be read. To this end, the text layout is systematically annotated using nine classes. 

    From a computer science view, a collection of successful approaches allows us to process modern documents
    \citep{blecher2023nougat, davis2022end}. For historical documents, large-scale data sets exist
    \citep{dell2024american} but are mostly focused on English language material
    set in Antiqua-like typefaces.
    For continental European languages, existing datasets are much smaller~
    \citep{abadie2022benchmark,nikolaidou2022survey,Kodym2021Page,clausner2015enp}

    Until more annotated data becomes available, the processing of historical continental European newspaper pages is,
    therefore, a low-resource task, highlighting the need for more data. While low-resource tasks are well-established
    in natural language processing \citep{adams2017cross,fadaee2017data,Heddrich2021Survey,Zoph2016Transfer}
    , low-resource settings remain under-explored in computer vision \citep{zhang2024low}
    . Historical German newspapers are interesting in this context due to their dense layout (see also Supplementary
    Figure \ref{fig:LayoutErrors}
    ) and the Fraktur font. Fraktur differs significantly from the Antiqua typefaces that dominate modern Western texts.
    To the contemporary eye, Fraktur letters appear dense, which also impacts layout recognition.
    Furthermore, in addition to the font, our dataset features the archaic 'long s' or '
            {\fontfamily{jkpvos}\selectfont{}s}
    ', which is no longer used today. The 'sz' or 'ß' is specific to the German language and also appears in the data.
    Historically, it emerged when the common combination '{\fontfamily{jkpvos}\selectfont{}s}z' merged into a single letter 'ß', unlike the 'long s' it still appears in contemporary texts. The aforementioned differences limit our ability to transfer existing solutions designed for modern documents or English-language
    historical newspapers. This motivates the collection of additional data.

\begin{figure}
        \centering
        \includestandalone[width=0.3\linewidth]{./figures/papers_per_year}
        \hspace{1cm}
        \includegraphics[width=0.225\linewidth]{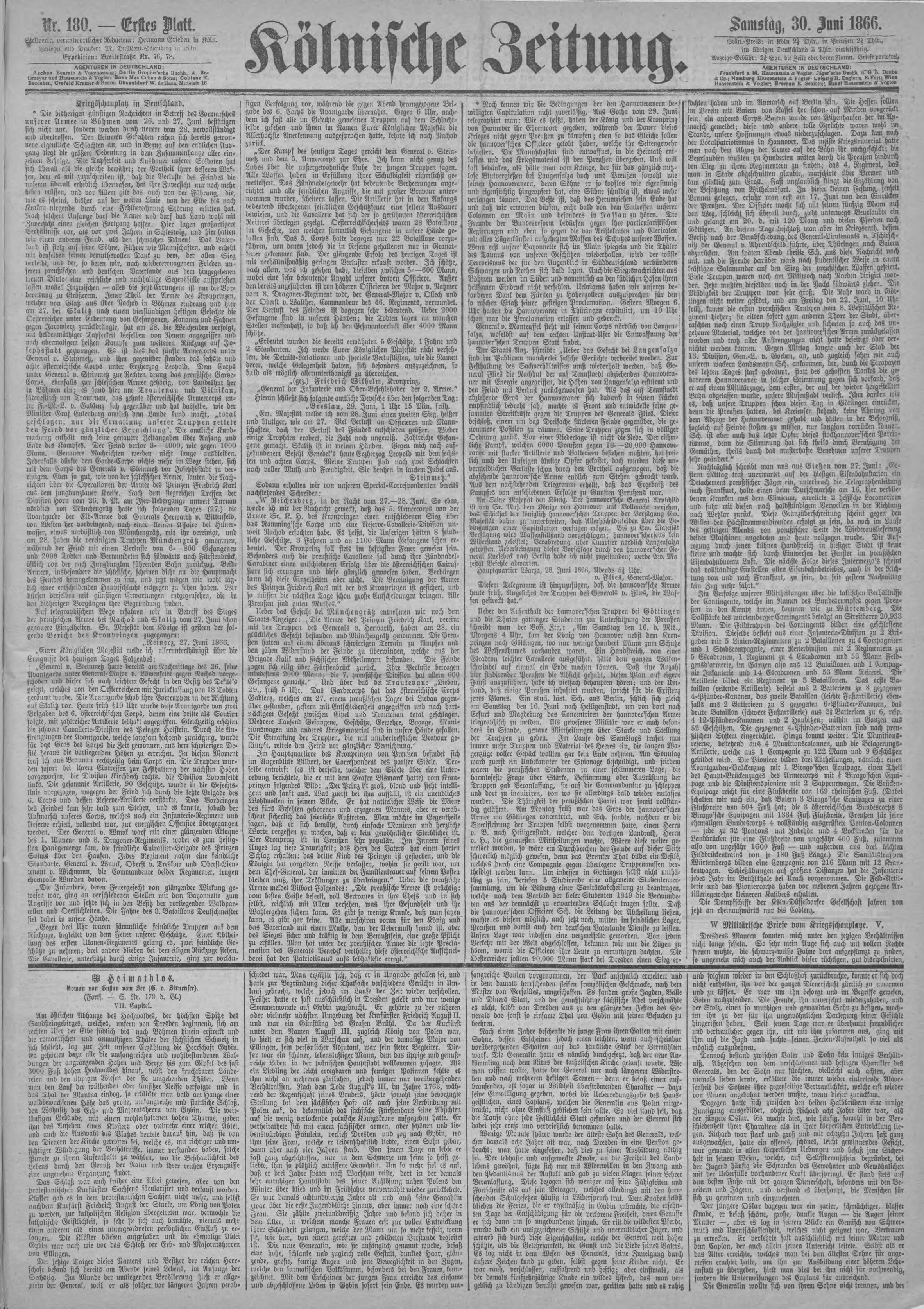}
        \includegraphics[width=0.225\linewidth]{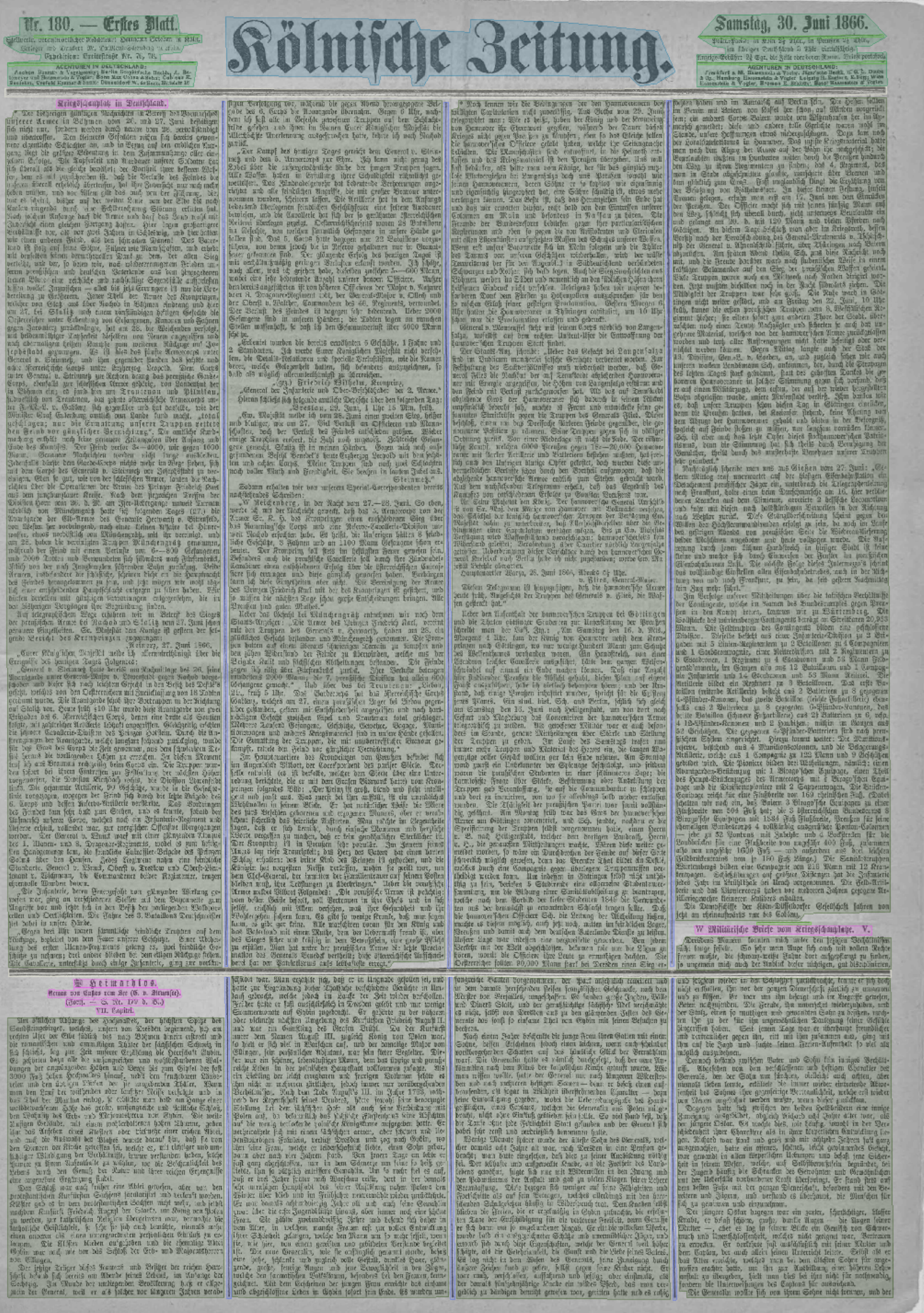}
        \caption
        {Left: Number of available digitized newspapers per year at \url{www.deutsche-digitale-bibliothek.de/newspaper}
        over time. Data from January 2024. Center: Front page of the \textit{Kölnische Zeitung} from the $1^{st}$ of January 1924. Right: Corresponding annotation to the page at the center of this figure.}
        \label{fig:papersperyear}
    \end{figure}

    The task of processing German newspapers is also highly relevant to historians.
    Especially in the 19th century, local communities, interest groups, and political parties created their own
    newspapers. The \textit{Deutsche Zeitungsportal}\footnote{https://www.deutsche-digitale-bibliothek.de/newspaper/}
    counts 698 German newspapers in 1780, this number rose to over 14,000 in 1860 and peaked at 50,848 papers in 1916 (see Figure \ref{fig:papersperyear}).

    Plenty of digitized pages allow researchers to systematically search for cultural values and historical change.
    Unfortunately, most of the pages available on such platforms contain either no or incorrect layout information. 
    With text lines being in disarray, the pre-processed text data from these pages cannot be analysed computationally. 
    Additionally, untrained modern human readers struggle with font differences, limiting the usefulness of unprocessed
    data to researchers lacking this specific skill. Thus, creating a pipeline capable of accurately processing this
    vast amount of data to a format readable to both a machine and a researcher without specific language and typeface
    skills is an important step in making these resources accessible. Furthermore, the availability of machine-readable
    newspaper archives is valuable to social scientists who recently started to use historical newspapers to track
    treatment variables or measure the impact of institutions or policies on social life \citep{Beach2023}.

    \begin{table}
        \centering
        \caption{Datasheet listing newspaper names, page counts, lines, words and polygon-regions. As layouts change in a paper's history, we include issues from different years. In such cases a paper appears more than once in the table. Newspapers that have been used as out-of-distribution test data only are marked with *}
        \label{tab:datasheet}
            \begin{tabular}{llrrrr}
                \toprule
                Year      & Newspaper                             & Pages & Lines   & Words     & Regions \\ \midrule
                1617      & Newspaper with no title, 30-Years war & 8     & 268     & 2,535      & 24     \\
                1626      & Newspaper with no title, 30-Years war & 8     & 276     & 2,618      & 28     \\
                1700      & Reichs Post Reuter                    & 8     & 350     & 2,991      & 42     \\
                1748      & Leipziger Zeitung                     & 4     & 292     & 2,051      & 56     \\
                1785      & Holzmindensches Wochenblatt           & 8     & 238     & 1,722      & 31     \\
                1785      & Schwäbischer Merkur                   & 11    & 1017    & 7,060      & 77     \\
                1813      & * Donau Zeitung                       & 4     & 304     & 1,706      & 47     \\
                1813      & Königlich privilegierte Stuttgarter Zeitung & 4 & 361   & 2,271      & 46     \\
                1834      & Fränkischer Kurier                    & 4     & 412     & 3,130      & 62     \\
                1849      & * Hamburgischer unpartheiischer Correspondent & 4 & 2,250 & 15,711   & 167    \\
                1851      & Ostpreussische Zeitung                & 4     & 1,183   & 9,538      & 126    \\
                1852-1888 & Special Issues Kölnische Zeitung      & 20    & 826     & 8,063      & 134    \\
                1856      & Der Bazar                             & 11    & 1,333   & 10,503     & 114    \\
                1857      & Berliner Börsen Zeitung               & 5     & 1,189   & 8,048      & 169    \\
                1866      & Bonner Zeitung                        & 4     & 1,400   & 10,364     & 176    \\
                1866      & Fränkischer Kurier                    & 6     & 1,814   & 12,375     & 302    \\
                1866      & Kölnische Zeitung                     & 420   & 248,295 & 2,137,972  & 17,050 \\
                1866      & Neue Berliner Musikzeitung            & 8     & 1,036   & 8,261      & 116    \\
                1866      & Pfälzer Zeitung                       & 4     & 1,080   & 7,666      & 127    \\
                1866      & Vossische Zeitung                     & 31    & 4,671   & 34,057     & 1,129  \\
                1866      & Weisseritz Zeitung                    & 8     & 832     & 5,795      & 114    \\
                1867      & Hannoverscher Courier                 & 4     & 1,946   & 13,905     & 226    \\
                1867      & Neue preussische Zeitung              & 4     & 3,179   & 23,067     & 203    \\
                1870      & Weisseritz Zeitung                    & 8     & 767     & 4,950      & 187    \\
                1871      & * Karlsruher Zeitung                  & 4     & 1,308   & 9,440      & 131    \\
                1889      & Mode und Haus, Illustrirte Kinderwelt & 8     & 342     & 2,976      & 66     \\
                1891      & Bonner Zeitung                        & 4     & 1,469   & 9,690      & 406    \\
                1898      & Dresdner Journal                      & 8     & 3,744   & 26,657     & 370    \\
                1917      & Muenchner Neue Nachrichten            & 6     & 3,421   & 19,292     & 621    \\
                1918      & * Darmstädter Zeitung                 & 4     & 1,388   & 9,309      & 207    \\
                1923      & Eibenstocker Tagblatt                 & 4     & 1,176   & 7,577      & 191    \\
                1924      & Kölnische Zeitung                     & 141   & 79,444  & 677,256    & 9,059  \\
                1928      & Dortmunder Genera-Anzeiger, Unterhaltungsblatt & 2 & 1,127 & 7,530   & 81     \\
                1929      & Rheinisches Volksblatt, Illustrierte Beilage & 4 & 554  & 3,796      & 125    \\
                1930      & Hildener Rundschau, Illustrierte Beilage & 8  & 726     & 4,617      & 185    \\
                1930      & * Lippische Zeitung, Volksblatt Illustrierte & 4 & 573  & 4,565      & 102    \\
                1933      & Vossische Zeitung                     & 4     & 1,051   & 8,071      & 154    \\  \midrule
                \multicolumn{2}{l}{Sum} & 801 & 371,642 & 3,127,135 & 32,451 \\ \bottomrule
            \end{tabular}
            \label{table:datasheet}
    \end{table}

    Additionally, the layout of German historical newspapers is often complex, consisting of several columns, multiple
    horizontal sections and up to 500 elements to annotate per page. To create this dataset, eleven student assistants
    with a background in history have spent a total of 2,700 hours annotating the layout of 801 pages. These include
    approx. 1,900 individually annotated advertisements that consist of approx. 5,700 polygon regions. We also provide
    ground truth text annotations, which are not as costly since we start from network-generated \ac{ocr}
    -output and correct errors. Overall, our dataset includes more than 30,000 layout polygon regions as well as more than
    370,000 text lines and approx. 3 million words. See Table~\ref{table:datasheet}
    for an overview of the dataset details.

    \begin{table}
    \centering
        \caption{Data-set sizes of this paper and related work in comparison.}
        \label{tab:data_set_comparison}
        \begin{tabular}{l c c} \toprule
        Dataset                                                 & Pages                                   \\ \midrule
        Chronicling Germany (ours)                              & \textbf{801}                            \\
        Europeana  \citep{clausner2015enp}                      & 528                                     \\
        News Eye Finnish     \citep{muehlberger_2021_5654858}   & 200                                     \\
        Reichs- und Staatsanzeiger  \citep{UBMannheimReich}     & 197                                     \\
        News Eye French \citep{muehlberger_2021_5654841}        & 183                                     \\
        Neue Züricher Zeitung \citep{clematide-stroebel-2019}   & 167                                     \\
        News Eye Austrian \citep{muehlberger_2021_5654907}      & 158                                     \\
        News Eye Competition \citep{michael_2021_4943582}       & 100                                     \\ \bottomrule
        \end{tabular}
    \end{table}

    This dataset is larger than the Europeana corpus \cite{clausner2015enp}
    with its 528 pages from European newspapers, the 197-page "Deutscher Reichsanzeiger und Preußischer Staatsanzeiger"
    \citep{UBMannheimReich} data set or the 167-page Neue Züriche Zeitung \citep{clematide-stroebel-2019, UBMannheimNZZ} corpus. Ground truth pages are compared in Table~\ref{tab:data_set_comparison}. Our dataset adds a significant number of new pages.

    Our dataset features sections and elements that are especially challenging for \ac{ocr}
    and baseline models. For example, advertisement pages mix large and small font sizes and include drop capitals, where
    the initial letter of an advertisement spans over multiple rows but is read as part of the first row. Both features
    are a challenge for the baseline detection task. Other challenges are fractions in stock exchange news and
    abbreviations in lists of casualties.\footnote{Our dataset includes pages from 1866, when the Austro-Prussian War was raging in the German Bund.}
    Overall, 83,8\% of our dataset scans fall into a range between 4500 and 5499 width times 6500 and 7499 height pixels.

    In summary, this paper makes the following contributions: (1) We introduce the \textit{Chronicling Germany}-dataset. Its 801 manually annotated high-resolution pages make it the largest German-language historic newspaper dataset to date (see Table~\ref{tab:datasheet} for dataset details). (2) We establish a baseline recognition pipeline for the layout detection, text-line
    recognition, and \ac{ocr}-tasks. (3) We verify generalization properties using the \ac{ood} part of our test set which contains 20 pages from five historic newspapers. We observe good generalization performance for OCR and layout tasks.
    We share processing code as well as the data via online repositories (see below).
    \begin{itemize}
        \item Data: \url{https://gitlab.uni-bonn.de/digital-history/Chronicling-Germany-Dataset}
        \item Code: \url{https://github.com/Digital-History-Bonn/Chronicling-Germany-Code}
    \end{itemize}

    \section{Related Work}\label{sec:related_work}

    Unfortunately, from a digital history perspective, many modern systems focus on recent data and suffer from poor
    performance in a historical setting.\footnote
    {For an example see Figure \ref{fig:LayoutErrors} in \ref{sec:suppl_additional_fig},
        alco compare \cite{Shen_2020_CVPR_Workshops}.} The current situation has led to a large body of \ac{ocr}
    error correction work \citep{carlson2023efficient}, highlighting the need for specialized data sets and software.
    \cite{Liebl2020FromNewsToData}, for example, combine existing open-source components for this task.

    Related datasets include the Europeana corpus~\citep{clausner2015enp}, the Deutsche Reichsanzeiger
    \citep{UBMannheimReich}, and the Neue Züricher Zeitung \citep{UBMannheimNZZ,clematide-stroebel-2019}. The Europeana dataset contains \textit{528} annotated pages from European sources. The Reichsanzeiger and the Neue Züricher Zeitung sets consist of \textit{197} and \textit{167} annotated pages, respectively, but have so far only been used for \ac{ocr} training but not in any layout training pipeline (cf. \ref{tab:data_set_comparison}). The layout annotation of these two projects is comparable to ours
    but less granular, and we have annotated considerably more pages.\footnote
    {We aim to combine those two datasets with Chronicling Germany in future work.}
    More recently \cite{dell2024american}, published perhaps the largest American historical newspaper dataset to date. Their dataset also includes layout
    annotations. Our work complements these existing datasets by additionally providing compatible annotations for German historical newspapers that differ significantly from other Western European and American newspapers.
    Furthermore, we annotate advertisements in detail, which significantly add to the complexity of the \ac{ocr}-task
    \citep{dell2024american}
    and are not annotated in the Reichsanzeiger and the Neue Züricher Zeitung. Advertisements are particularly
    interesting to scholars of economic history who are interested in labour markets, for example. Globally, the
    Historical Japanese Dataset \citep{Shen_2020_CVPR_Workshops}
    , a collection of over 2,000 annotated pages of complex layouts from the Japanese Who is Who of 1953, is notable.

    \subsection{Common processing pipeline elements}

    \paragraph{Layout Segmentation} is a longstanding task in document processing. For example, dhSegment
    \citep{oliveira2018dhsegment} proposes a UNet structure based on the popular ResNet50 architecture
    \citep{he2016deep}. As described by \cite{ronneberger2015u}, the network features a contracting and an expanding part. The contracting subnetwork uses ResNet50 as an encoder,
    and an additional expansive subnetwork produces segmentation maps at the resolution of the original input.
    Transformer-based solutions trained on modern documents are available for similar tasks \citep{davis2022end}. However, \acp{cnn} are cheaper to run \citep{dell2024american} and require less training data, making them a budget-friendly solution.

    \paragraph{Baseline-detection} or text-line detection means finding the straight line that connects the base points from each letter. Early work employed quadratic splines for this task \citep{smith2007overview}. Modern solutions often employ architectures devised for segmentation or object detection tasks. \cite{Kodym2021Page} for example, choose a U-Net.
    Object detection pipelines are alternatively used instead of baseline detection; e.g., \cite{dell2024american} work with YOLOv8.
    Following \cite{Kodym2021Page}, we employ a U-Net to detect text baselines and use them to generate line polygons. Our annotations are consistent with the Europeana-corpus from \cite{clausner2015enp} and the work from \cite{UBMannheimNZZ,UBMannheimReich} that also features Fraktur letters. This design choice allows us to combine our datasets in future work.

    \paragraph{\acf{ocr}} is an important tool in digital history. \cite{Liebl2020FromNewsToData}, successfully work with a topological feature extraction step followed by a classifier as described by \cite{smith2007overview} for the digitization of the \textit{Berliner Börsen Zeitung}. Following \cite{breuel2007announcing}, \cite{Kiessling_The_Kraken_OCR_2022} uses a \ac{rnn} based system. \cite{dell2024american} apply the contrastive learning approach presented by \cite{carlson2023efficient}. Using a vision encoder, characters are projected into a metric space. The system works because patches containing
    the same character will cluster together.

    \section{The Chronicling Germany Dataset}\label{sec:dataset}
    Our Dataset contains 801 pages from historic German newspapers, mostly from 1866, specifically from the period of the Austro-Prussian War. Of these 801 pages, 15 pages contain only advertisements with approx. 1,900 individual advertisement blocks. The backbone of the dataset is the \textit{Kölnische Zeitung}, a large regional newspaper from Western Germany. Additionally we include newspaper pages from before and after the German Empire and various German regions (See Table~\ref{table:datasheet}). Overall, we consider the dataset to be a very good representation of the different layout styles of historical German newspapers. (For a more detailed description and justification of the composition of the dataset, see \ref{sec:dataset_description}).
    
    We split our data into train, validation, and test datasets (Table~\ref{table:label-histogramm}), where the test dataset consists of \acf{id} and \acf{ood}. \ac{ood} pages are taken from five German newspapers that are not present in the training or validation dataset (Table~\ref{table:datasheet}). The train and validation data splits only contain \ac{id} data. 

    Polygons placed by our expert human annotators capture the layout for each page.\footnote
    {The annotation process is documented in \ref{sec:dataset_description},
        the annotation guidlines are reported in \ref{sec:suppl_annotation_guide}.}
    All annotations are stored in PAGE-XML files. The polygons capture different text-region types. Subclasses can exist within these. Each region type has a unique XML tag: \texttt{TextRegion, SeparatorRegion, TableRegion} and \texttt{GraphicRegion}.
    Graphic regions are always assigned the class \texttt{image}. Within text regions, we include the following classes:
    \texttt{paragraph,}
    \texttt{header,}
    \texttt{heading,}
    \texttt{caption,}
    \texttt{inverted\_text}.
    Within table regions, the only possible subclass is
    \texttt{table}.
    To facilitate correct reading order detection, we introduce the separator subclass
    \texttt{separator\_vertical,} and
    \texttt{separator\_horizontal}.
    \begin{table}
        \caption
        {Dataset split (left), test data is divided into \acf{id}- and \acf{ood}-data. 
         The right hand side shows label distribution-percentages per pixel.
        }
        \begin{tabular}{l c c c}
            \toprule
            & pages & polygons & lines \\ \midrule
            train          & 651   & 27,162    & 324,399    \\
            validation     & 50    & 1,784     & 15,024    \\
            test \acs{id}   & 80    & 2,851     & 26,396    \\
            test \acs{ood}  & 20   & 654     & 5823    \\\midrule
            sum            & 801   & 32,451    & 371,642    \\\bottomrule
        \end{tabular}
        \begin{tabular}{c l l}
            \toprule
            label & class                & frequency \\ \midrule
            0       & background           & 34.11\%   \\
            1       & caption              & 0.77\%    \\
            2       & table                & 3.10\%    \\
            3       & paragraph            & 58.50\%   \\
            4       & heading              & 1.11\%    \\
            5       & header               & 0.74\%    \\
            6       & separator vertical   & 0.68\%    \\
            7       & separator horizontal & 0.63\%    \\
            8       & image                & 0.32\%   \\
            9       & inverted text        & 0.02\%   \\ \bottomrule
        \end{tabular}
        \label{table:label-histogramm}
    \end{table}
    Vertical separators highlight different columns of a page. Horizontal separators split the page into sections and
    are relevant for the reading order if they span over multiple columns. Otherwise, they are found at the beginning of
    a new article or between caption or header elements. The header category covers the newspaper's name, which appears
    at the top of the front pages. To the left and right of the newspaper name, historical newspapers often have smaller
    blocks with additional information, such as the name of the editor-in-chief, the publication date, or the
    subscription price. These polygons are annotated as captions. Polygons that cover paragraphs, headlines, and tables
    are annotated, respectively. See Figure~\ref{fig:papersperyear}
    for an annotation sample. Overall, the dataset includes 32,451 polygon regions.
    
    We primarily use a combination of the classes described above to annotate the historic advertisements. We have
    decided not to introduce new classes to avoid confounding the model's training. This applies, in particular, to the
    separator classes. Therefore, we use the classes \texttt{separator\_vertical} and \texttt{separator\_horizontal}
    for the annotation of separator regions around individual advertisements. Advertisements tend to use text blocks
    with bigger fonts. To be consistent with our annotations, we mark these as \texttt{heading}. For the same reason, the normal-sized text is annotated as \texttt{paragraph}. Additionally, we include the classes \texttt{inverted\_text} and graphic elements as \texttt{image}. These are present, especially in the advertisement pages, as well as in pages from 1924.
    Table~\ref{table:label-histogramm} illustrates this numerically. The two classes \texttt{inverted\_text} and
    \texttt{image} are only present in a subset of the data, which explains its low share of pixels overall.

    Regions of each page have a reading order number assigned to them. These numbers are assigned automatically and not
    corrected manually. Reading order is not the main scope of this dataset. Automatic assignment leads to satisfactory
    results for most pages. For advertisement pages, however, it does not. Yet, advertisements don't need a meaningful
    reading order, as they are comprised of elements that are independent of each other.


    In addition to the layout data, we include transcribed text divided into text lines. In our dataset, each text line
    is comprised of a polygon, which contains all characters, as well as a baseline and the corresponding text.
    Baselines and text transcriptions are generated automatically and then corrected by expert annotators. Partially the generation was done using the pipeline proposed by
    \cite{Kodym2021Page}
    , more recent additions to the dataset have been generated using our own pipeline. Line polygons and baselines are only corrected when there are significant
    mistakes. This is especially the case within the advertisement pages, where some initial letters of advertisements
    span over more than one line. Correct drop capital detection is challenging for current text-line detectors. Overall, our dataset includes 371,642 text lines. The transcription follows the OCR-D guidelines, level 2
    \citep{Mangei2024GTGuide}. This means the text is transcribed in a visual style, preserving, for example, the archaic 'long s' or '
            {\fontfamily{jkpvos}\selectfont{}s}'. For a complete discussion, see supplemental section
    \ref{sec:suppl_annotation_guide}.

    \section{Experiments and results}\label{sec:results}

    \begin{figure}
        \centering
        \includestandalone[width=1\linewidth]{figures/pipeline/pipeline}
        \caption{Flow chart of the entire prediction pipeline. The layout detection,
            text-line inference and \acf{ocr}-tasks use separate networks each. The output is machine-readable and can
            be processed further. For example in a machine translation step.}
        \label{fig:pipeline}
    \end{figure}

    \paragraph{Data:} All experiments work with fixed train, validation and test splits as outlined in Table~\ref{table:label-histogramm}.

    \paragraph{Pipeline:} Figure~\ref{fig:pipeline}
    presents a pipeline overview. Overall, we employ two U-Nets for layout recognition and text-line detection and, finally, a \ac{lstm} network for \ac{ocr}. The pixel-wise layout inference is converted into polygons during the post-processing step. We use targets like
    \cite{Kodym2021Page} for training the baseline U-Net.
    The model recognizes baselines, ascender-, descender-, and end-points, which are converted into line regions and baselines during post-processing. The post-processing code is an adapted version from \cite{Kodym2021Page}.
    Contrary to their approach, we use the layout regions from the previous step to cut out parts of the image and
    identify all baselines for each region. These baselines are then used as input for the \ac{lstm} \ac{ocr}
    model and the original image. 

    \subsection{Layout-Segmentation}\label{sec:layout-seg-exp}
    \paragraph{Training:} Our layout segmentation setup follows \cite{oliveira2018dhsegment}.
    For layout training, all pages are scaled down by a factor of 0.5 and split into 512 by 512-pixel crops. Cropping
    leads to 39,072 training crops overall.
    During training, we work with 32 crops per batch per graphics card. The
    training runs on a node with four \acp{gpu}. Consequently, the effective batch size is 128, with 305 training steps per epoch.
    Initially, optimization of the contracting network part can start from pre-trained ImageNet weights, while
    optimization of the expanding path has to start from scratch. The expanding subnetwork starts with the encoding from the contracting network and produces a segmentation output at the input resolution. To improve generalization, input crops are augmented using rotation, mirroring, Gaussian blurring, and randomly erasing rectangular regions. An AdamW-Optimizer \citep{Loshchilov2017AdamW} trains this network with a learning rate of 0.0001, with a weight decay parameter of 0.001 for 50 Epochs in total, while using early stopping to save the best model.
    We initialize the ResNet-50 encoder using ImagNet weights before training. We use only in distribution data for training and validation.


    \begin{table}
        \centering
        \caption
        {Layout detection test set results. This table lists F1 Score values for all individual classes. N/A values in columns (3) and (5) are due to differing annotations between our approach and \cite{dell2024american}.  In columns (2) and (4)
            we report our model results for \acf{id} + \acf{ood} and \acf{id} only data. (Out of distribution only results for layout in Table~\ref{table:layout-results-ood-only})}
        \begin{tabular}{l c c c c}
            \toprule
            & \multicolumn{2}{c}{F1 Score (\acs{id} + \acs{ood})} & \multicolumn{2}{c}{F1 Score (\acs{id} only)}\\
            \cmidrule{2-3} \cmidrule{4-5}
            class                   & ours                  & \citeauthor{dell2024american} & ours                  & \citeauthor{dell2024american}\\ \midrule
            background              & 0.96 $\pm$ 0.01      & 0.79                         & 0.93 $\pm$ 0.03      & 0.79   \\
            caption                 & 0.63 $\pm$ 0.10      & N/A                           & 0.85 $\pm$ 0.01      & N/A     \\
            table                   & 0.90 $\pm$ 0.02      & 0.43                         & 0.90 $\pm$ 0.03      & 0.45   \\
            paragraph               & 0.97 $\pm$ 0.01      & 0.88                         & 0.95 $\pm$ 0.03      & 0.89   \\
            heading                 & 0.67 $\pm$ 0.04      & 0.46                         & 0.69 $\pm$ 0.03      & 0.46   \\
            header                  & 0.80 $\pm$ 0.03      & 0.23                         & 0.80 $\pm$ 0.05      & 0.26   \\
            separator vertical      & 0.72 $\pm$ 0.02      & N/A                           & 0.72 $\pm$ 0.03      & N/A     \\
            separator horizontal    & 0.74 $\pm$ 0.01      & N/A                           & 0.73 $\pm$ 0.02      & N/A     \\
            image                   & 0.70 $\pm$ 0.06      & N/A                           & 0.74 $\pm$ 0.06      & N/A     \\
            inverted text           & 0.16 $\pm$ 0.02      & N/A                           & 0.27 $\pm$ 0.14      & N/A     \\ \bottomrule
        \end{tabular}
        \label{table:layout-results}
    \end{table}

    \paragraph{Results:} Table~\ref{table:layout-results} column two lists network performance on the test dataset, and column four lists performance on the id test data only. We compute F1 Score values for all individual classes on pixel level. Generally, we find good performance with both datasets.  Table~\ref{table:layout-results-ood-only} additionally shows results for \acs{ood} only data. Overall, the results show that rare classes, especially \texttt{image} and \texttt{inverted\_text}, are a challenge, while the \texttt{paragraph} class shows very good performance and generalization. Most noticeably, our pipeline displays much better performance for the \texttt{caption} and \texttt{header} classes for \ac{id} in contrast to \ac{ood} data (Table~\ref{table:full-layout-results}). We suspect that this behavior is caused by specific caption patterns. In \ac{ood} data, the caption pattern is often unusual compared to \ac{id} data, which creates this effect. Figure~\ref{fig:target_and_predction} and Figure~\ref{fig:target_and_predction_2}  present newspaper pages from our test set with layout ground truth and prediction samples side by side.

    We also compare our model results to the pipeline developed by \cite{dell2024american}. To his end, we run their pipeline on our test data. Columns three and five of table ~\ref{table:layout-results} list the results.
    Overall, our pipeline performs better on the comparable classes of our test dataset than \cite{dell2024american}.
    However, we do not fine-tune their model on our training data, and there are significant differences between the Chronicling Germany dataset and the American Stories dataset that distort the comparison. Dell et al. work with a YOLOv8 object detection model for layout recognition. We have not been able to fine-tune their pipeline. Instead, we evaluate another YOLOv8 model that has been fine-tuned on our dataset. Table~\ref{table:layout_yolo} shows full results of the fine-tuned YOLOv8. The most significant difference between Dell et al. and our data is the more detailed headline annotation of the Chronicling Germany dataset. \footnote{Segmentation of headlines is of particular interest for historical research, as they allow for the layout-based identification of and differentiation between individual articles.}
    \cite{dell2024american} seems to assign the headline class less frequently to headlines that do not stand out clearly. This leads to a significant amount of headline regions from the Chronicling Germany test-set to be classified as paragraphs by the Dell-pipeline. Furthermore, the Chronicling Germany dataset includes annotated separator regions, while American Stories does not.\footnote
    {In the case of the Kölnische Zeitung, identifying horizontal page-spanning separators is of crucial importance for
    the reading order, as they act like a page break. This means the reader should not continue reading down the current column but go back up to the next one.} Moreover, \cite{dell2024american} treats the entire header of a newspaper front page as one class, while Chronicling Germany differentiates between the header itself and the captions that are typically left and right of the header.

    \subsection{Baseline Detection}
    \textbf{Training}: Following \cite{Kodym2021Page}
    we train an U-Net for the text-baseline prediction task. The raw input image as well as ground truth baselines serve as starting points for the optimization.
    The training process minimizes a joint text-line and text-block detection objective as introduced by
    \cite{Kodym2021Page}. We run an AdamW-optimizer \citep{Loshchilov2017AdamW} with a learning rate of 0.0001 and a batch size of 16. During training, inputs are randomly cropped to 256 by 256 images.
    To improve the robustness of the resulting network, the input pipeline includes color jitter, Gaussian blur, random
    grayscale, random scaling, and Gaussian blur perturbations during training.

    \paragraph{Results:} We measure precision, recall, and F1 score (see Table~\ref{table:baseline_results}
    ). Generally, we observe values around 0.9. These observations are in line with \cite{Kodym2021Page},
     who observe similar numbers on the cBAD2019 dataset \citep{diem2017cbad}. \cite{dell2024american}
    do not provide baseline data, instead opting for extracting the text for each region as a whole employing a Yolov8
    architecture. Since we do not annotate the ground-truth text boxes, there is no adequate way to compare the two
    pipelines for this specific task. The historic newspaper community either works with baseline detection or direct
    text object detection pipelines. We decided to follow \cite{Kodym2021Page}
    and employ a U-Net to detect text baselines. This is consistent with other European projects like the
    Europeana-corpus from \cite{clausner2015enp}
    and the work from UB-Mannheim that also features Fraktur letters. This choice allows combining these European
    datasets in future work, and is a key design choice, since we aim to boost performance in the Fraktur-subset of
    historical newspapers.

    \begin{table}
        \centering
        \caption{Baseline detection test set results. We measure performance in precision,
            recall and F1 score. Detected lines are matched with ground truth lines and are considered a true positive
            if the predicted line has an IoU score of more than 0.7 compared to the corresponding ground truth line.
            Results are averaged over all test pages.}
        \begin{tabular}{l c c c}
            \toprule
            Model                       & precision         & recall            & F1 score          \\ \midrule
             UNet                        & 0.938 $\pm$ 0.007 &  0.897 $\pm$ 0.013 & 0.917 $\pm$ 0.010 \\
        \bottomrule
        \end{tabular}
        \label{table:baseline_results}
    \end{table}

    \subsection{\acf{ocr}}
    \textbf{Training:} Based on the Kraken-OCR-engine \citep{Kiessling_The_Kraken_OCR_2022} we train a \ac{lstm}-cell for the \ac{ocr}
    -task and employ baselines to extract individual line polygons. Alongside the annotations, which our human domain
    experts have checked, these serve as input and ground truth pairs. Adam \citep{Kingma2015Adam}
    optimizes the network with a learning rate of 0.001. Optimization runs for eight epochs with a batch size of 32
    sequences. We use early stopping to prevent the models from overfitting and include pixel-dropout, blur, rotation, and see-through-like augmentations during training to improve
    generalization.
    Additionally, we train and evaluate the OCR-transformer proposed by \cite{Kodym2021Page} as part of their pero-application with AdamW optimizer \citep{Loshchilov2017AdamW} and a learning rate of 0.0001, using the same augmentations.\\
    \begin{table}
        \centering
        \caption
        {
        \acf{ocr} test set results. Levenshtein distance per character appears in the first column. We compute the percentage
        of completely error-free lines for each model. The second column lists these results. Finally, we consider a line to have many errors if we observe a Levenshtein distance of more than 0.1 per character. We report the percentage of many error lines in the final column. We list the mean and standard deviation for multiple seeds.}
        \begin{tabular}{l l l l l}
          \toprule
          Model                                                 &    Data               & Levenshtein-Distance  & fully correct [\%] & many errors [\%] \\
          \midrule
          \multirow{2}{*}{LSTM (UBM \citeyear{Kamlah2024ocr})}  & \ac{id} + \ac{ood}    & 0.029                           & 47.1                        &  6.1                    \\
                                                                & \ac{id} only          & 0.029                           & 48.5                        &  6.0                    \\
                                                                & \ac{ood} only         & 0.030                           & 40.8                        &  6.4                    \\  \midrule[.02em]
          \multirow{2}{*}{LSTM finetuned (ours) }               & \ac{id} + \ac{ood}    & 0.025 $\pm$ 0.002               & 67.3 $\pm$ 0.6              &  4.3 $\pm$ 0.2          \\
                                                                & \ac{id} only          & 0.026 $\pm$ 0.002               & 70.1 $\pm$ 0.5              &  4.4 $\pm$ 0.2          \\
                                                                & \ac{ood} only         & 0.018 $\pm$ 0.001               & 59.2 $\pm$ 1.5              &  4.1 $\pm$ 0.3          \\  \midrule[.02em]
          \multirow{2}{*}{LSTM (ours) }                         & \ac{id} + \ac{ood}    & 0.022 $\pm$ 0.002               & 68.2 $\pm$ 0.7              &  4.0 $\pm$ 0.2          \\
                                                                & \ac{id} only          & 0.023 $\pm$ 0.002               & 70.4 $\pm$ 0.6              &  4.0 $\pm$ 0.3          \\
                                                                & \ac{ood} only         & 0.019 $\pm$ 0.001               & 58.4 $\pm$ 1.7              &  4.1 $\pm$ 0.2          \\  \midrule[.02em]
          \multirow{2}{*}{Transformer (ours) }                  & \ac{id} + \ac{ood}    & 0.028 $\pm$ 0.001               & 64.1 $\pm$ 0.4              &  8.6 $\pm$ 0.3          \\
                                                                & \ac{id} only          & 0.021 $\pm$ 0.001               & 69.5 $\pm$ 0.3              &  5.3 $\pm$ 0.2          \\
                                                                & \ac{ood} only         & 0.065 $\pm$ 0.003               & 39.5 $\pm$ 0.8              & 23.7 $\pm$ 1.0          \\
          \bottomrule
        \end{tabular}
        \label{table:ocr_results}
    \end{table}

    \paragraph{Results:} Compared to the model trained by the Universitäts­bibliothek Mannheim     \citep{Kamlah2024ocr}, we observe slightly improved performance when finetuning their model as well as training a new model from scratch (Table~\ref{table:ocr_results}). Overall, we achieve a Levenshtein distance of 0.022 per character, which corresponds to 97.8\% of all characters being correctly recognized. We will refer to this character recognition rate as the score.
    
    The corresponding model achieves a perfect score on 68.2\% of all text lines, and 96\% of all text lines achieve a score of at least 90\%. (see: Table~\ref{table:ocr_results}
    ).  Overall, our OCR-transformer achieves a score similar to the original \citep{Kamlah2024ocr} Kraken model. Our LSTM cell performs slightly better. We notice, however, that the OCR-transformer produces the worst results on \ac{ood} test data. At the same time, it performs best on \ac{id} data. We suppose that this is a sign of overfitting on the \ac{id} data. The Antiqua-pretrained OCR model from \cite{dell2024american}
    does not generalize well to the Fraktur texts. For their pipeline, we observe an average Levenshtein distance of 0.58
    on the test set (not included in Table~\ref{table:ocr_results}).\footnote
    {Please note that we cannot fine-tune the OCR engine proposed by \cite{dell2024american} on our Fraktur data because
    of differences in the text detection step (see above).}

    \subsection{Overall pipeline performance}
    So far, we have evaluated all components individually using ground truth inputs from previous steps. Next, we additionally
    run the complete pipeline on the test set (Table~\ref{table:pipeline_ocr_results}). We choose the best model for each component, according to the validation set (50 pages),
    to use in our pipeline. Then, we evaluate the resulting
    transcription with our ground truth. All predicted and ground truth lines are matched based on the intersection over
    union of the corresponding text lines. Lines without a match are paired with an empty string. Our pipeline
    achieves an overall Levenshtein distance per character of $0.07$ (93\% score) across the entire test dataset. The significant difference to the \ac{ocr}-only evaluation in Table~\ref{table:ocr_results} is due to errors in the layout prediction step. These errors cause prediction text lines to not be matchable with the ground truth lines.
    Table ~\ref{table:pipeline_reduced_ocr_results} compares results of the original pipeline with a modified version, where different text regions are merged and unmatched lines are ignored. The modified pipeline evaluation results in a Levenshtein distance of 0.032 (96.8\% score), similar to the OCR-only evaluation (Table ~\ref{table:ocr_results}).

    \section{Pipeline-Generalization}
    \begin{figure}
        \centering
        \includegraphics[height=5cm]{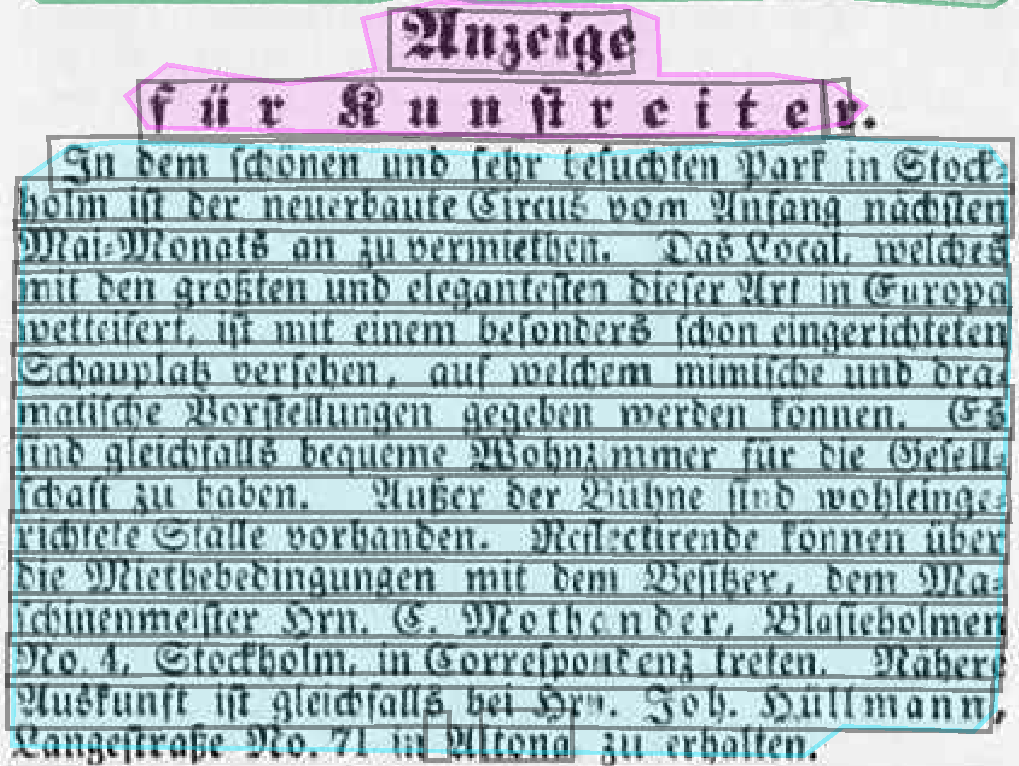}
        \resizebox{!}{5cm}{
            \begin{tikzpicture}
                \node[rectangle,draw, align=center] {Anzeige\\
                	für Kun{\fontfamily{jkpvos}\selectfont{}s}treite\\
                	In dem {\fontfamily{jkpvos}\selectfont{}s}chönen und {\fontfamily{jkpvos}\selectfont{}s}ehr be{\fontfamily{jkpvos}\selectfont{}s}uchten Park in Stock-\\
                	holm i{\fontfamily{jkpvos}\selectfont{}s}t der neuerbaute Circus vom Anfang näch{\fontfamily{jkpvos}\selectfont{}s}ten\\
                	Mai-Monats an zu vermiethen. Das Local, welches\\
                	mit den großten und elegante{\fontfamily{jkpvos}\selectfont{}s}ten die{\fontfamily{jkpvos}\selectfont{}s}er Art in Europa\\
                	wetteifert, i{\fontfamily{jkpvos}\selectfont{}s}t mit einem be{\fontfamily{jkpvos}\selectfont{}s}onders {\fontfamily{jkpvos}\selectfont{}s}chon eingerichteten\\
                	
                	Schauplatz ver{\fontfamily{jkpvos}\selectfont{}s}ehen, auf welchem mimi{\fontfamily{jkpvos}\selectfont{}s}che und dra-\\
                	mati{\fontfamily{jkpvos}\selectfont{}s}che Vor{\fontfamily{jkpvos}\selectfont{}s}tellungen gegeben werden konnen. Es\\
                	{\fontfamily{jkpvos}\selectfont{}s}ind gleichfalls bequeme Wohnzimmer für die Ge{\fontfamily{jkpvos}\selectfont{}s}ell-\\
                	{\fontfamily{jkpvos}\selectfont{}s}cha{\fontfamily{jkpvos}\selectfont{}s}t zu haben. Außer der Bühne {\fontfamily{jkpvos}\selectfont{}s}ind wohleinge-\\
                	richtete Stalle vorhanden. Reflectirende können über\\
                	die Miethebedingungen mit dem Be{\fontfamily{jkpvos}\selectfont{}s}itzer, dem Ma-\\
                	{\fontfamily{jkpvos}\selectfont{}s}chinenmei{\fontfamily{jkpvos}\selectfont{}s}ter Hrn. C. Mothender, Bla{\fontfamily{jkpvos}\selectfont{}s}iebolmen\\
                	No. 4. Stockholm, in Corre{\fontfamily{jkpvos}\selectfont{}s}pondenz treten. Nähere\\
                	Auskunft i{\fontfamily{jkpvos}\selectfont{}s}t gleichfalls bei Hrn. Joh. Hüllmann\\
                	ltona};
            \end{tikzpicture}
        }
        \caption
        {Generalization test set sample image. This figure shows a page element with detected baselines on the left. The
        right side presents the automatically created transcription.}
        \label{fig:generalise_example}
    \end{figure}

    \begin{table}
        \centering
        \caption
        {\acf{ocr} test set results for running the entire pipeline on our dataset. Levenshtein distance per character appears in the first column. We compute the percentage
        of completely error-free lines for each model. The second column lists these results. Finally, we consider a line to have many errors if we observe a Levenshtein distance of more than 0.1 per character. We report the percentage of many error lines in the final column. All predicted and ground truth lines are matched based on the intersection over
        the minimum of the corresponding text lines. }
        \begin{tabular}{l c c c c}
        \toprule
        Model            &    Data                  & Levenshtein-Distance & fully correct [\%] & many errors [\%] \\
        \midrule
        \multirow{3}{*}{Full Pipeline}  & \ac{id} + \ac{ood}  & 0.070     & 53.4         & 21.1     \\
                    &   \ac{id} only  & 0.065     & 55.7         & 19.7    \\
                    &   \ac{ood} only  & 0.080     & 43.3         & 27.6   \\
        \bottomrule
        \end{tabular}
        \label{table:pipeline_ocr_results}
    \end{table}

    \paragraph{Test-Data:} To verify generalization, our test dataset contains 20 \acf{ood} pages from different
    papers and time periods (Table~\ref{table:datasheet} and Table~\ref{table:label-histogramm}).

    \paragraph{Inference:} In addition to the full test set evaluation, we run the entire pipeline on the \ac{ood} data only
	(Table~\ref{table:pipeline_ocr_results}). We measure a Levenshtein distance per
    character of $0.08$ (92\% score). When ignoring unmatched lines and merging different text regions into one label, the results significantly improve to 0.038 (96.2\% score) (Table ~\ref{table:pipeline_reduced_ocr_results}). Figure~\ref{fig:generalise_example}
    presents an \ac{ood} example taken from a 1849 issue of the Hamburger Correspondent. The sample is an advertisement for a circus in Stockholm that is available for rent.
    The advertisement describes the quarters and stables which belong to the circus. It includes the name and address of the owner, to allow interested parties to reach out for additional information. The names and addresses contain several abbreviations, which are a challenge for \ac{ocr}. In this example, most abbreviations are recognized correctly.  Linguistically, the sample is close enough to modern German to be machine-translated.

    \section{Conclusion and future work}\label{sec:conclusion}
    This work introduces the \textit{Chronicling Germany}
    -dataset, the currently largest dataset of German language historical newspaper pages. In addition to the dataset, it presents a neural network-based processing baseline with test-set \ac{ocr}-accuracy results.
    Our paper creates a starting point for researchers who wish to improve historical newspaper processing pipelines or are looking for a low-resource computer vision challenge.
    To create the dataset, history students spent 2,700 hours annotating the layout. The dataset's 801 pages make it the largest fully annotated collection of historic German newspaper pages to date. The dataset includes 1,900 individually annotated advertisements. Furthermore, the \acf{ood} part of our test set includes 20 pages selected from historic newspapers that are not part of the training or validation set.
    We verify baseline pipeline performance on the \ac{ood} pages. By following the OCR-D annotation guidelines \citep{Mangei2024GTGuide}, we ensure our annotations' compatibility with concurrent and future work.

    \subsection*{Acknowledgments}
    We thank Jan Göttfert, Charlotte Lüttschwager, and Ida Wenzel for helping annotate the data. We furthermore want to thank Paul Asmuth, Judith Eifler, Daniel Göttlich and Moritz von Kolzenberg, and Robin Weiden for supporting the project.

    This project was funded by the University of Bonn's TRA1 (Mathematics, Modelling, and Simulation of
    Complex Systems) and  TRA4 (Individuals, Institutions, and Societies)
    as part of the Excellence Strategy of the German federal and state governments.
    In particular, we would like to extend our gratitude to Dr. Daniel Minge and Johanna Tix for helping fund our dataset collection efforts.
    Furthermore, research was supported by the Bundesministerium für Bildung und Forschung (BMBF) via its  "BNTrAInee"
    (16DHBK1022) and "WestAI" (01IS22094A) projects. 
    
    The authors gratefully acknowledge the Gauss Centre for Supercomputing e.V. (www.gauss-centre.eu) for funding this project by providing computing time through the John von Neumann Institute for Computing (NIC) on the GCS Supercomputer JUWELS at Jülich Supercomputing Centre (JSC).

    \bibliography{literature}

    \newpage
    \appendix

    \section{Supplementary Material}
    
    \subsection{Acronyms}
    \printacronyms[heading=None]

    \subsection{Machine learning is important for the study of history}\label{sec:hist_motivation}
    Figure \ref{fig:papersperyear} illustrates the breakthrough of the newspaper industry in the 19th century.\footnote
    {Note that the Deutsche Zeitungsportal does not collect all historical newspapers. There is probably a selection
    bias towards more prominent outlets with extended publication periods. However, on the whole,
        figure \ref{fig:papersperyear} should reflect the development of the newspaper market in Germany quite well. }
    While the number of newspapers listed in the \textit{Deutsche Zeitungsportal}
    grew at a rate of 2.9 percent p.a. in the first two-thirds of the 19th century, the increase rose to 3.4 percent
    p.a. after the foundation of the German Empire. There were three main reasons for this increase: firstly, the
    literacy of the population increased over the century. Secondly, considerable technological advances made it easier
    to produce a newspaper. Thirdly, state control of newspapers declined from the middle of the century. A significant
    milestone was the Press Act of 1874, which finally abolished censorship (although some restrictions remained so that
    even after 1874, there was no complete freedom of the press in the German Empire). Nevertheless, no later than the
    last third of the 19th century, a mass market for print media had emerged in Germany, which was served by many
    newspapers whose content and political orientation were very heterogeneous.

    Historical newspapers contain a wealth of information about past societies. They provide information about the
    spatial occurrence of events, about contemporary perceptions of social and economic change, and allow tracing of
    cultural change. \cite{Blevins2014}, for example, uses the mentioning of place names in the \textit{Houston Post}
    to draw a mental map of the Nation around 1900. Measured by the mention of place names, the region west of Houston
    was deeply rooted in the newspaper and its readership. The East Coast and the Midwest were also present in the
    imagination of contemporaries. However, the Southwest, the Northwest, and California hardly appear on this mental
    map. Based on the newspaper's coverage, one could argue that readers of the \textit{Houston Post}
    around 1900 were barely aware of the Nation as a geographical entity. In economic history, historical newspapers
    have recently been used to identify treatments or measure variables of interest. \cite{Beach2023}
    give an overview of the recent use of historical newspaper data in economic history. An interesting recent example
    is \cite{Ferrara2024}, who used digitized newspaper archives to measure a county's exposure to the boll weevil around 1900. The boll
    weevil is a pest of cotton that hit the American South between 1892 to 1922. The pest reduced cotton production and,
    consequently, hastened social changes in the primarily Black rural communities, like the fertility transition and
    higher investment in education.

    Even though newspaper portals are an essential source for historians and other disciplines interested in history,
    such as economics, their potential has not yet been fully realized \citep{Beach2023}
    . Firstly, researchers have so far mainly used US-American portals. The reason for this bias may be these portals
    have been established longer than in other regions of the world. Secondly, the mass utilization of newspaper data is
    often limited to a keyword search, which usually only covers the entire page and does not discriminate between
    articles. Therefore, the joint occurrence of two or more search terms is recorded for the page, not the article, and
    information retrieval is thus still very imprecise \citep{Oberbichler2021Topic}. Thirdly, the text cannot always be downloaded easily, which makes further processing by researchers more
    difficult. On the other hand, the image files of individual newspaper pages are easy to obtain via the portals. Deep
    learning algorithms that recognize the layout of a newspaper page and capture the text at the article level,
    therefore, promise great benefits for historical research. The \textit{Chronicling Germany}
    data set presented here, comes with layout annotations for every page. It is intended to stimulate the further
    development of deep learning algorithms and to promote the increased use of non-American newspaper portals.

    In addition to more accurate and straightforward information retrieval, downloadable article-level data will also
    allow scholars of history to apply advanced NLP-methods in the future, including document and text embedding
    techniques and fine-tuning large language models to 19th-century German.

    \subsection{Additional figures}\label{sec:suppl_additional_fig}

    \begin{figure}[H]
        \centering
        \includegraphics[width=0.32\linewidth]{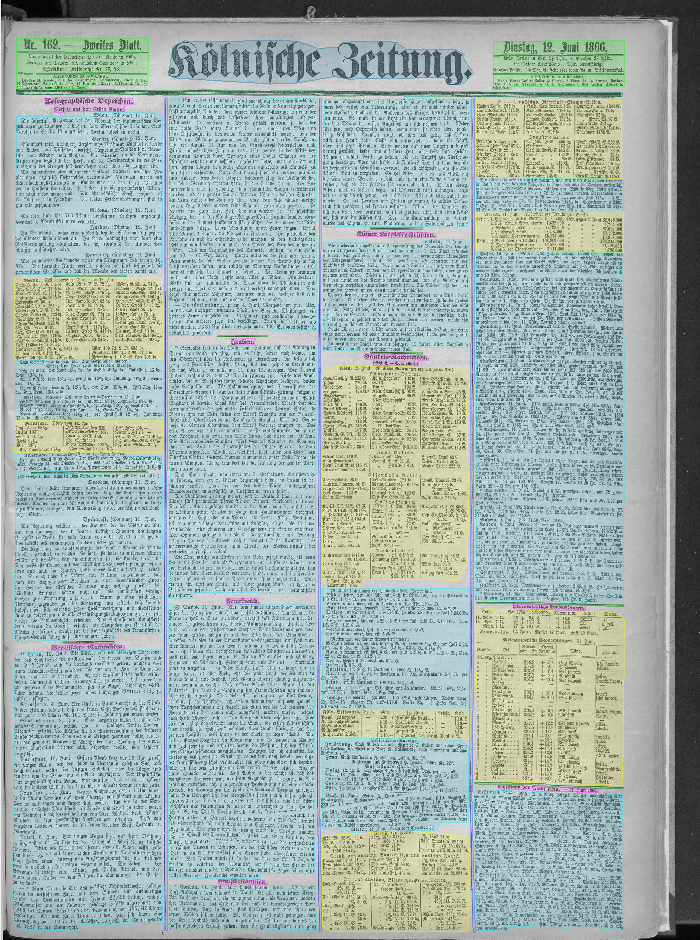}
        \includegraphics[width=0.32\linewidth]{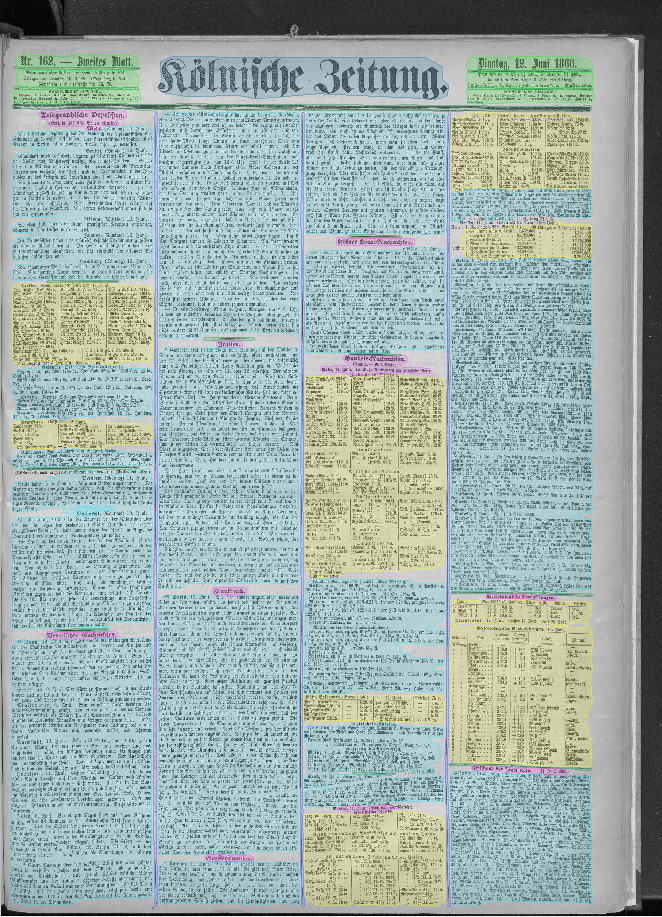}
        \includestandalone[width=0.25\linewidth]{figures/legend/legend}
        \caption
        {Target labels on the left and segmentation prediction on the right. Newspaper page of the Kölnsche Zeitung from the \ac{id} test set.}
        \label{fig:target_and_predction}
    \end{figure}

        \begin{figure}[H]
        \centering
        \includegraphics[width=0.32\linewidth]{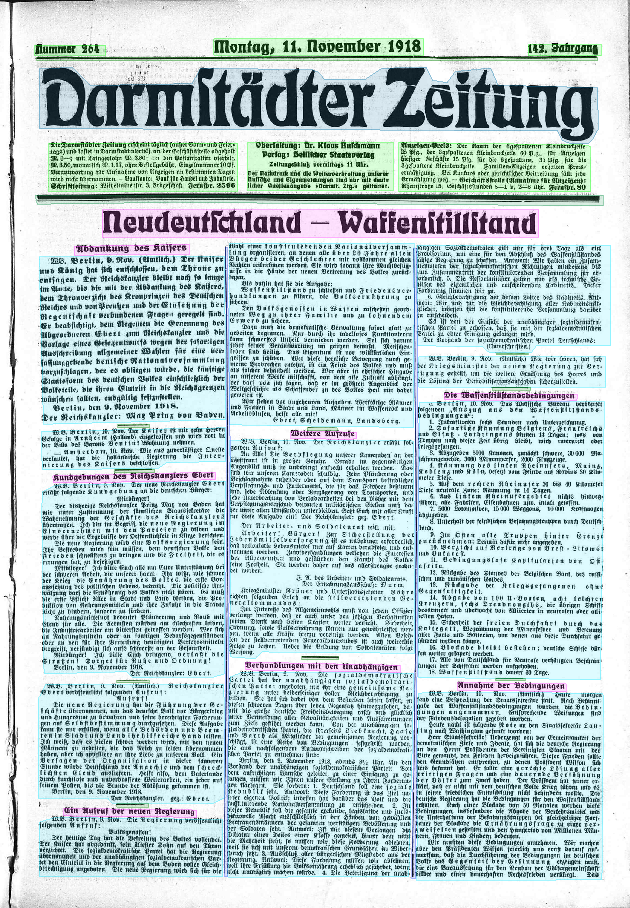}
        \includegraphics[width=0.32\linewidth]{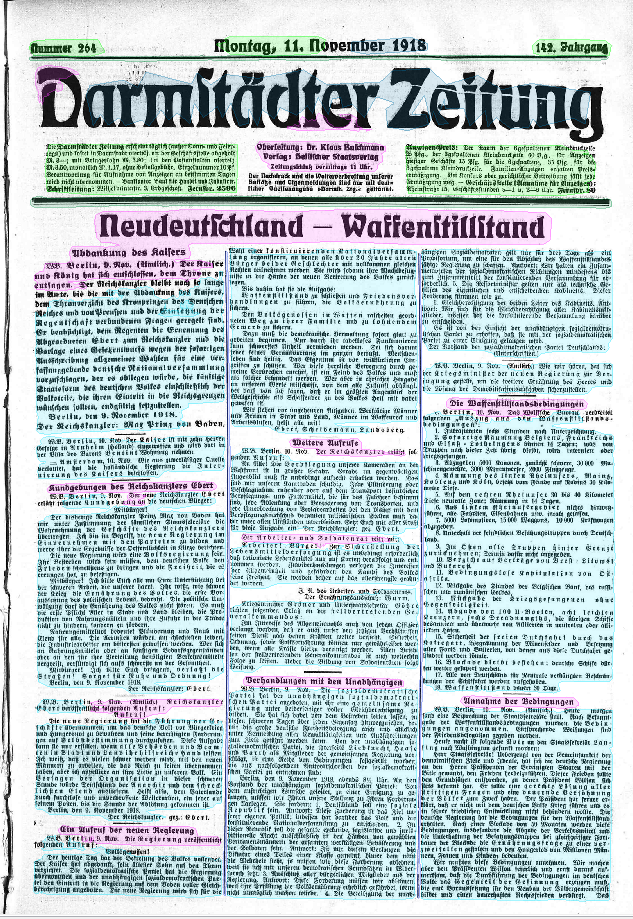}
        \includestandalone[width=0.25\linewidth]{figures/legend/legend}
        \caption
        {Target labels on the left and segmentation prediction on the right. Newspaper page of the Darmstädter Zeitung from the \ac{ood} test set.}
        \label{fig:target_and_predction_2}
    \end{figure}

    \begin{figure}[H]
        \includegraphics[width=\textwidth]{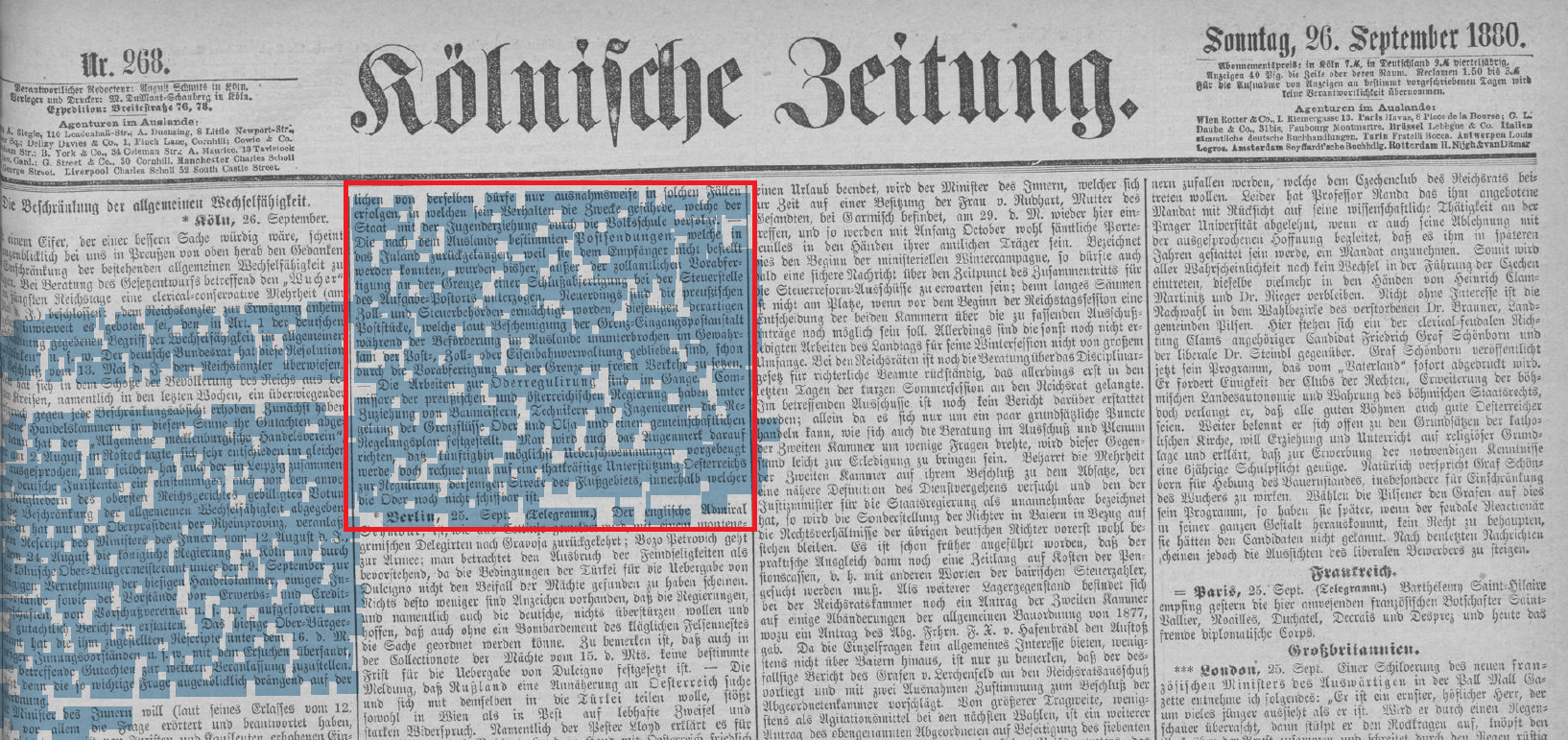}
        \caption
        {Layout recognition error in the \textit{Kölnische Zeitung}. A researcher tried to select text in the first
        column on the very left but the column layout is not understood correctly. This page was published on September
        26, 1880. Digital versions are available at the \textit{zeit.punkt NRW} website. Layout recognition and
        transcription generated by Transkribus.
        }
        \label{fig:LayoutErrors}
    \end{figure}

    \subsection{Dataset description}\label{sec:dataset_description}

    \subsubsection{Construction of the dataset}
\begin{figure}
        \centering
        \includegraphics[width=.9\linewidth]{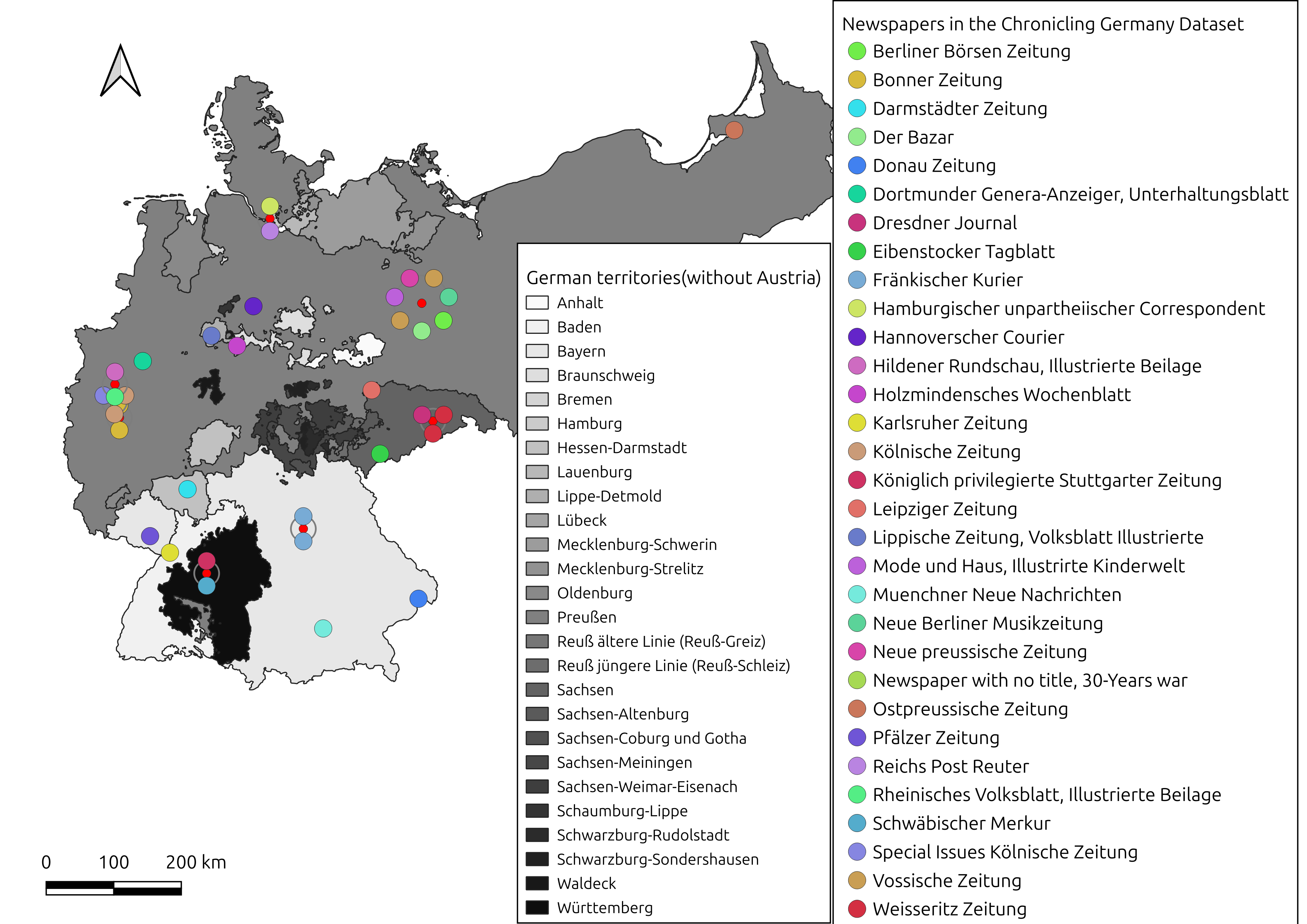}
        \caption{Map of historic Germany from 1867 with labeled regions and regions of
        newspaper origins beyond the Kölnische Zeitung.}
        \label{fig:newspaper-map}
    \end{figure}

    The Chronicling Germany dataset includes 30 newspapers from all over Germany (Figure: \ref{fig:newspaper-map}). Most of the data (581 pages) is from the Kölnische Zeitung but we have added 220 pages from other newspapers
    covering the time period 1617-1933 (Table \ref{table:datasheet}). Currently, these 30 newspapers are completely in the test set. However, in an updated version we will also include different newspapers into the training pipeline. Overall, we have annotated 801 pages including over 3 million
    words and ca. 32,000 region polygons.
    Aside from availability and representation, we selected these newspapers for the following reasons: Our focus is on
    1866, the year of the Austro-Prussian War. Aside from its historical importance as the second of the three
    unification wars and the decisive turning point towards the "Kleindeutsche Lösung", this year gives certain
    advantages to make our data more diverse: During this year, most newspapers across Germany reported lists of the
    fallen, missing and deserted as well as reports on military careers of officers. These are usually printed in a
    considerably different layout, thus diversifying our data. Additionally, most states - particularly during wartime -
    obliged newspapers in their territory to publish "Öffentliche Bekanntmachungen" or official notices. In 1866, all
    states handled this differently, resulting in more diverse newspaper layouts across Germany (compared to 1871). Also
    , focusing on this year allows users of our dataset to evaluate separately how well a model generalizes to different
    newspapers of the same time and how well it generalizes to newspapers from other decades. When no newspapers from
    1866 were available, we sometimes included an issue from 1867. The different newspapers have been chosen to maximize
    variation between them. We include newspapers from various regions of (past) Germany, like Berlin, Eastern Prussia,
    the Rhineland, Lower Germany (Hannover), Bavaria, The Palatinate, and Saxony. We also take care to include larger
    national as well as regional newspapers, and newspapers with a special non-political interest, like the Neue Berliner
    Musikzeitung (New Berlin Musics Paper) and Der Bazar (a paper on "women's topics" - mostly written by men). We
    dedicated extra attention to the Vossische Zeitung, because it is one of the most-read newspapers of its age and -
    due to its bad printing quality - it is particularly difficult for layout detection and OCR. The amount of pages per
    newspaper varies, since we include full issues of each newspaper, regardless of their length. This is done to ensure
    that the entire diversity in layout and font across different sections of the newspaper is represented in the
    dataset. The years from 1924 onwards constitutes a natural end for the dataset, since German newspapers gradually
    started using latina fonts instead of Fraktur during that period.

    \subsubsection{Annotation process}
    The annotation process follows the annotation guidlenes in \ref{sec:suppl_annotation_guide}. A human domain expert carried out all layout annotations that were then cross-checked by another human domain
    expert.
    Annotating and correcting text is extremely time-consuming. For the moment an automatic transcription is checked and corrected by a single human domain expert. Currently, we have corrected 446 pages in of the Kölnische Zeitung, and 112 pages in the generalization part of the dataset. To improve quality further we will run a second correction round, where all lines will again by proofread by different annotators.

    \subsection{Annotation Guidelines}\label{sec:suppl_annotation_guide}

\input{./guidelines/guide_text}

    \newpage
    \subsection{Further Results}\label{sec:suppl_further_results}
    \subsubsection{Modified Pipeline Results}
    \begin{table}[htbp]
   	\centering
   	\caption
   	{\acf{ocr} test set results for running the entire pipeline on our dataset. We compare the original pipeline with a modified version, where we apply layout merging that combines different text regions and therefore  labels all text regions as paragraph. All predicted and ground truth lines are matched based on intersection over union. The modified pipeline is evaluated in a way that ignores unmatched text lines.}
    \begin{tabular}{l c c c c}
    	\toprule
    	Model            &    Data                  & Levenshtein-Distance & fully correct [\%] & many errors [\%] \\
    	\midrule
        \multirow{3}{*}{Original}  & \ac{id} + \ac{ood}  & 0.070     & 53.4         & 21.1     \\
		&   \ac{id} only  & 0.065     & 55.7         & 19.7    \\
		&   \ac{ood} only  & 0.080     & 43.3         & 27.6  \\\midrule[.02em]
    	\multirow{3}{*}{Modified}  & \ac{id} + \ac{ood}  & 0.032     & 63.4         & 6.8     \\
    	&   \ac{id} only  &0.027     & 65.2         & 6.1    \\
    	&   \ac{ood} only  & 0.038     & 56.9         & 7.9   \\
    	\bottomrule
    \end{tabular}
    \label{table:pipeline_reduced_ocr_results}
    \end{table}

    \subsubsection{Additional Layout Results}
        \begin{table}[htbp]
    	\centering
    	\caption
    	{Full layout detection test set results for our pipeline. This table lists F1 Score values for all individual classes. As the \ac{ood} data does not contain the \texttt{inverted\_text} class, it can not be evaluated on \ac{ood} data.}
    	\begin{tabular}{l c c c}
    		\toprule
    		class & F1 Score (\acs{id} + \acs{ood}) & F1 Score (\acs{id} only) & F1 Score (\acs{ood} only)\\ \midrule
    		background           & 0.96 $\pm$ 0.004 & 0.93 $\pm$ 0.030 &  0.89 $\pm$ 0.022    \\
    		caption              & 0.63 $\pm$ 0.097 & 0.85 $\pm$ 0.010 &  0.39 $\pm$ 0.132    \\
    		table                & 0.90 $\pm$ 0.016 & 0.90 $\pm$ 0.027 &  0.67 $\pm$ 0.036    \\
    		paragraph            & 0.97 $\pm$ 0.005 & 0.95 $\pm$ 0.027 &  0.90 $\pm$ 0.025    \\
    		heading              & 0.67 $\pm$ 0.037 & 0.69 $\pm$ 0.025 &  0.50 $\pm$ 0.052    \\
    		header               & 0.80 $\pm$ 0.025 & 0.80 $\pm$ 0.047 &  0.40 $\pm$ 0.085    \\
    		separator vertical   & 0.72 $\pm$ 0.024 & 0.72 $\pm$ 0.027 &  0.58 $\pm$ 0.028    \\
    		separator horizontal & 0.74 $\pm$ 0.014 & 0.73 $\pm$ 0.016 &  0.57 $\pm$ 0.043    \\
    		image                & 0.70 $\pm$ 0.064 & 0.74 $\pm$ 0.063 &  0.38 $\pm$ 0.121    \\
    		inverted text        & 0.16 $\pm$ 0.017 & 0.27 $\pm$ 0.136 &  N/A    \\ \bottomrule
    	\end{tabular}
    	\label{table:full-layout-results}
    \end{table}

        \begin{table}[htbp]
        \centering
        \caption
        {Layout detection results on the out of distribution test set. This table lists F1 Score values for all individual classes. N/A values for \cite{dell2024american} are due to annotation differences. As the \ac{ood} data does not contain the \texttt{inverted\_text} class, it cannot be evaluated in this table. Since YOLOv8 detects bounding boxes, we do not require seperator detection. (Main results for layout in Table~\ref{table:layout-results})}
        \label{tab:layout_id_ood}
        \begin{tabular}{l c c c}
            \toprule
            &  \multicolumn{3}{c}{F1 Score (\acs{ood} only)}\\
            \cmidrule{2-4}
            class                     & ours                  & YOLOv8                     & \citeauthor{dell2024american}\\ \midrule
            background                & 0.89 $\pm$ 0.022      & 0.90 $\pm$ 0.001          & 0.82   \\
            caption                   & 0.39 $\pm$ 0.132      & 0.62 $\pm$ 0.092          & N/A    \\
            table                     & 0.67 $\pm$ 0.036      & 0.24 $\pm$ 0.098          & 0.16   \\
            paragraph                 & 0.90 $\pm$ 0.025      & 0.91 $\pm$ 0.012          & 0.84   \\
            heading                   & 0.50 $\pm$ 0.052      & 0.63 $\pm$ 0.020          & 0.43   \\
            header                    & 0.40 $\pm$ 0.085      & 0.62 $\pm$ 0.033          & 0.07   \\
            separator vertical        & 0.58 $\pm$ 0.028      & N/A                        & N/A    \\
            separator horizontal      & 0.57 $\pm$ 0.043      & N/A                        & N/A    \\
            image                     & 0.38 $\pm$ 0.121      & 0.00 $\pm$ 0.000          & N/A    \\
            inverted text             & N/A                   & N/A                        & N/A    \\ \bottomrule
        \end{tabular}
        \label{table:layout-results-ood-only}
    \end{table}

        \begin{table}[htbp]
        \centering
        \caption
        {Layout detection results. This table lists F1 Score values for all individual classes form a YOLOv8 detection model trained on our dataset. Since YOLOv8 detects bounding boxes, we do not require seperator detection.}
        \label{tab:layout_yolo}
        \begin{tabular}{l c c c c c c}
            \toprule
            class & F1 Score (\acs{id} + \acs{ood}) & F1 Score (\acs{id} only) &  F1 Score (\acs{ood} only)\\
            \midrule
            background           & 0.90 $\pm$ 0.003       & 0.90 $\pm$ 0.003      &  0.90 $\pm$ 0.001  \\
            caption              & 0.79 $\pm$ 0.135       & 0.80 $\pm$ 0.144      &  0.62 $\pm$ 0.092  \\
            table                & 0.75 $\pm$ 0.020       & 0.81 $\pm$ 0.027      &  0.24 $\pm$ 0.098  \\
            paragraph            & 0.94 $\pm$ 0.003       & 0.94 $\pm$ 0.002      &  0.91 $\pm$ 0.012  \\
            heading              & 0.72 $\pm$ 0.013       & 0.73 $\pm$ 0.015      &  0.63 $\pm$ 0.020  \\
            header               & 0.78 $\pm$ 0.023       & 0.80 $\pm$ 0.031      &  0.62 $\pm$ 0.033  \\
            image                & 0.03 $\pm$ 0.003       & 0.03 $\pm$ 0.004      &  0.00 $\pm$ 0.000  \\
            inverted text        & 0.00 $\pm$ 0.000       & 0.00 $\pm$ 0.000      &  N/A                \\ \bottomrule
        \end{tabular}
        \label{table:layout_yolo}
    \end{table}

  \subsection{Project Limitations and Future Work}\label{sec:limitations}
    \subsubsection{\ac{ocr}}
    \label{sec:limitations:antiqua}
    This dataset contains newspaper pages set in Fraktur-letters. The font is very different from
    modern fonts. The 'long s' or '{\fontfamily{jkpvos}\selectfont{}s}
    ', for example, is completely foreign to modern eyes. While our generalization dataset also includes four pages in
    Antiqua font, which have been predicted with sufficient accuracy, networks trained exclusively on our dataset are
    not likely to outperform more specialized networks on modern newspaper pages.

	\subsubsection{Layout Limitations}
\begin{figure}[H]
	\raggedleft
	\includegraphics[width=0.6\linewidth]{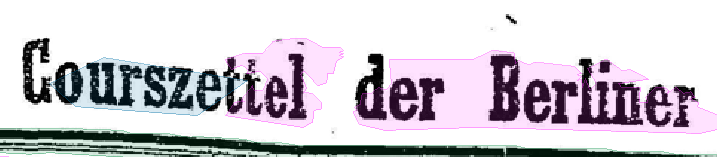}
	\centering
	\includegraphics[width=0.6\linewidth]{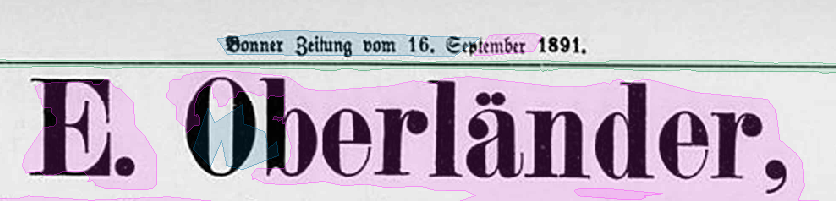}
	\includestandalone[width=0.25\linewidth]{figures/legend/legend_heading_header}
	\caption
	{Layout Prediction Examples for the challenge of differentiating headlines and header, which both have a higher font size than usual text.}
	\label{fig:headline_challenge}
	\end{figure}

	The full layout results (Table~\ref{table:pipeline_ocr_results}) show the continued challenge of differentiating between different text regions, especially in \ac{ood} data. This is due to strongly varying layout in different newspapers and the lack of sufficiently many and diverse examples in rare classes within our dataset. For the \texttt{caption} \texttt{header} classes, we observe a drastic drop in performance from \ac{id} to \ac{ood} data. One common error is, to confuse \texttt{header} and \texttt{headline} classes, as shown in Figure~\ref{fig:headline_challenge}.  For \texttt{caption} and \texttt{header} the limited receptive field of our layout model (\cite{oliveira2018dhsegment}) is an issue, as they are mostly detectable by considering the overall position on the newspaper page. We aim to improve this by either employing a CNN-based model with a much higher receptive field or using a detection model. Table~\ref{table:layout-results-ood-only} contains \ac{ood} only results for our layout model, the finetuned YOLOv8, and the \citeauthor{dell2024american} layout model. The finetuned YOLOv8 shows better generalization performance than our model on \texttt{header}, \texttt{heading}, and \texttt{caption} classes. However, the results for \texttt{table} are much worse than our model. Overall, we aim to further expand our dataset in the future, specifically increasing rare class examples.

\begin{figure}[H]
	\centering
	\includegraphics[width=0.32\linewidth]{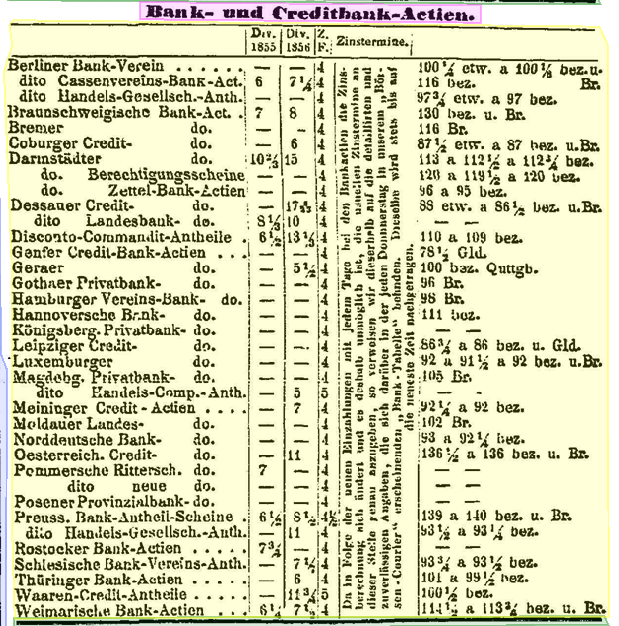}
	\includegraphics[width=0.32\linewidth]{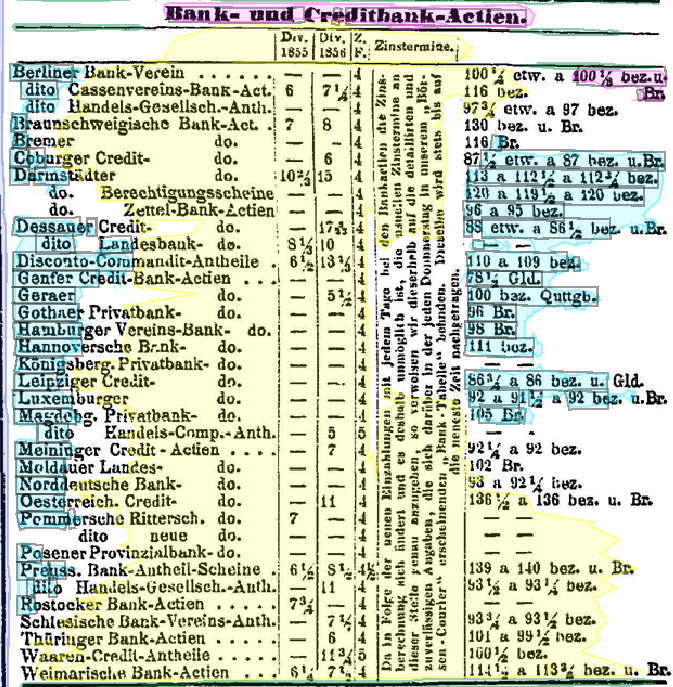}
	\includestandalone[width=0.25\linewidth]{figures/legend/legend}
	\caption
	{Target labels on the left and segmentation prediction on the right. Challenging table example from the \ac{id} test set.}
	\label{fig:table_challenge_1}
	\end{figure}

\begin{figure}[H]
	\centering
	\includegraphics[width=0.32\linewidth]{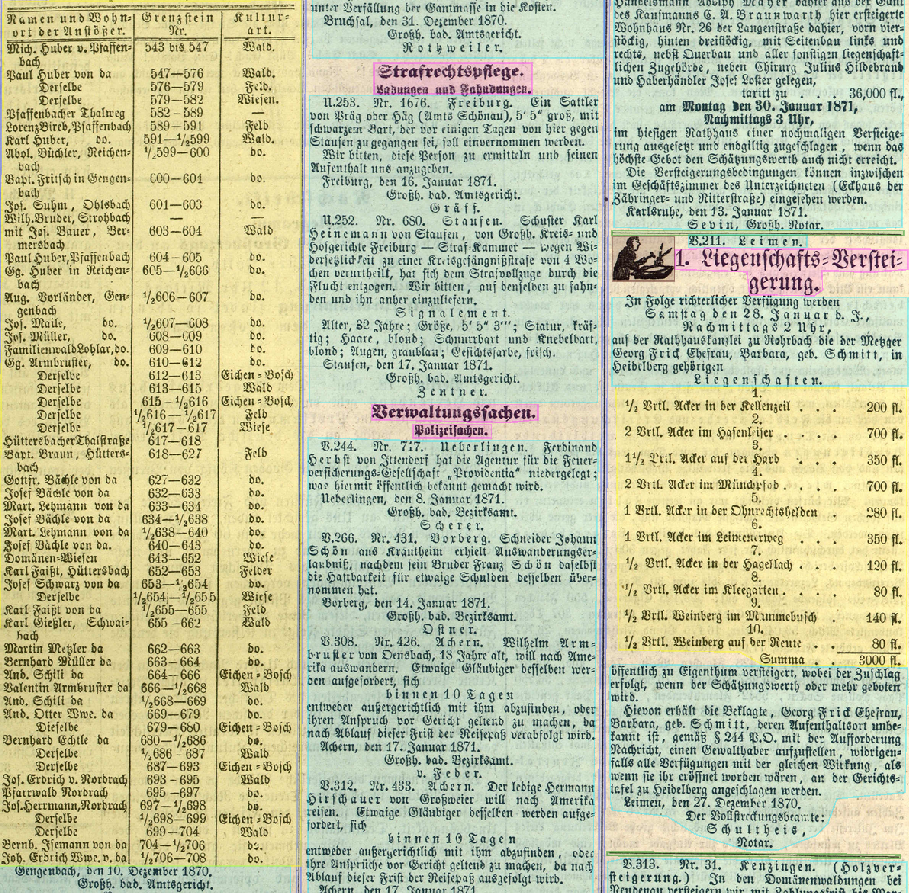}
\includegraphics[width=0.32\linewidth]{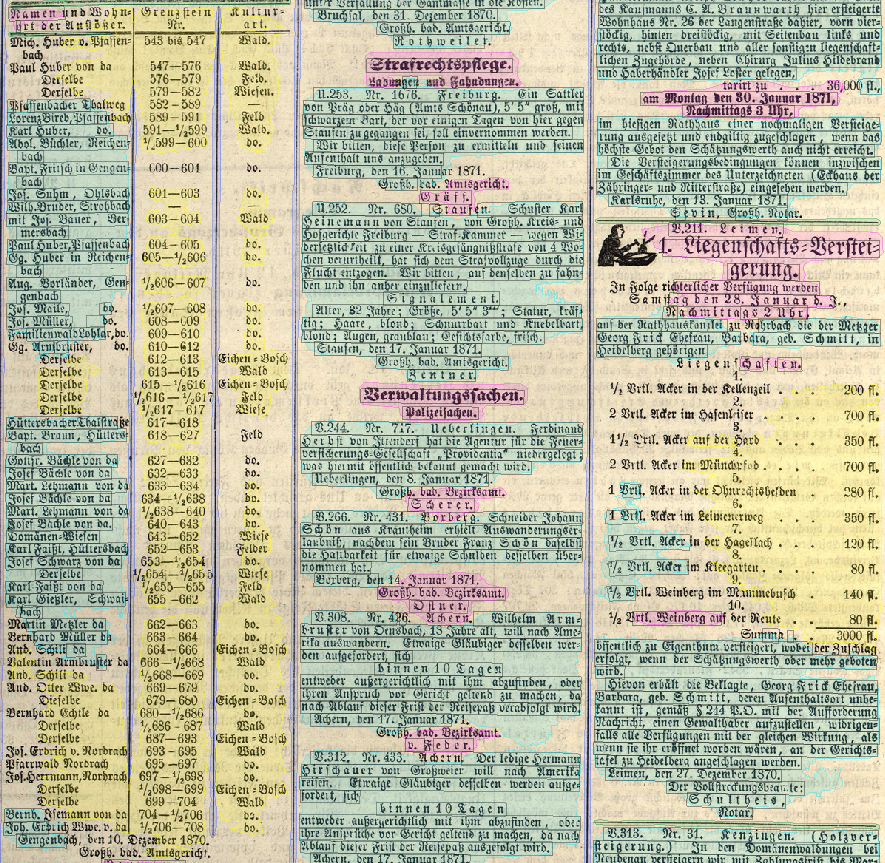}
	\includestandalone[width=0.25\linewidth]{figures/legend/legend}
	\caption
	{Target labels on the left and segmentation prediction on the right. Challenging table example from the \ac{ood} test set.}
	\label{fig:table_challenge_2}
	\end{figure}

\begin{figure}[H]
	\centering
	\includegraphics[width=0.32\linewidth]{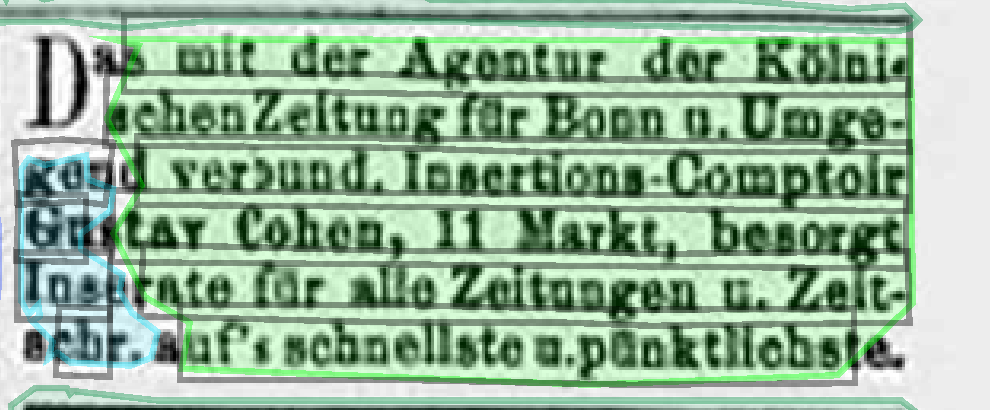}
	\includegraphics[width=0.32\linewidth]{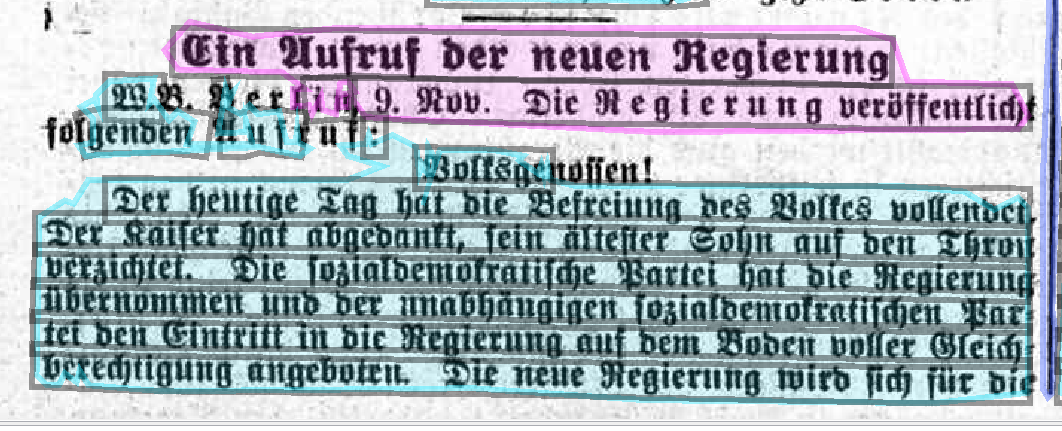}
	\includestandalone[width=0.25\linewidth]{figures/legend/legend_caption_heading_paragraph}
	\caption
	{Examples for split text lines due to layout errors.}
	\label{fig:split_lines}
\end{figure}

	Layout errors commonly cause text lines to be split into different text region classes, as shown in Figure~\ref{fig:split_lines}. Therefore, the pipeline will generate separate text line polygons for one ground truth text line. Another issue are tables, or part thereof, that are being mistreated as text regions, as shown in Figure~\ref{fig:table_challenge_1} and Figure~\ref{fig:table_challenge_2}. In the ground truth data tables do not have line polygons or text annotations. When the pipeline produces text lines in ground truth tables, those lines cannot be matched.

	This impacts the matching of text lines during evaluation and has a big effect on the overall pipeline performance. We aim to change the architecture of our pipeline in a way, that prevents text lines from being split by layout borders. Ideally, the split text lines in the left example of Figure~\ref{fig:split_lines} would be considered to be about 80\% caption and 20\% paragraph instead of being split at that region border. However, some layout information has to be considered when detecting baselines as they cannot be drawn from one column to another, ignoring column borders or even separators as shown in Figure~\ref{fig:LayoutErrors}. This might be achievable by splitting the layout segmentation task up in a basic text region detection task and a segmentation task, that differentiates between different text regions. Baseline detection has to be independent of the latter and dependent on the former.

	\subsubsection{Baseline Limitations}
\begin{figure}[H]
	\raggedleft
	\includegraphics[width=0.6\linewidth]{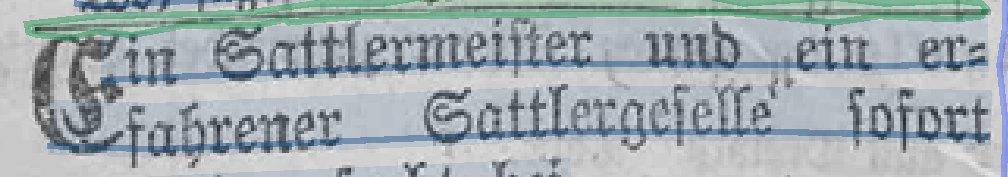}
	\centering
	\includegraphics[width=0.6\linewidth]{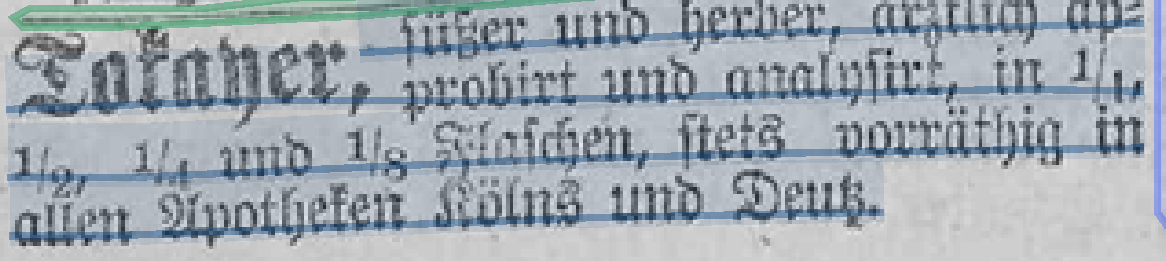}
	\caption
	{Examples for baseline errors caused by changing font sizes within text lines. On top is a drop capital that is the first character of the first line. The lower example is an entire word written in a bigger font. The sentence continues on the upper text line to the right of it.}
	\label{fig:drop_capital_challenge}
	\end{figure}

	Figure~\ref{fig:drop_capital_challenge} shows examples of baseline errors due to changing font size within text lines. Such examples occur almost exclusively in advertisements, therefore being very rare in our data. Our baseline detection model fails in correctly connecting baselines in such cases. We aim to add additional data containing drop capitals and font changes.

	\subsubsection{Article Separation}
	Article Separation is of great importance to enable further processing of transcribed pages. The main challenge lies in articles that span over page breaks and require the consideration of neighbouring pages. It is possible for multiple articles to span over the same page break. It has to be considered, that some newspaper pages are digitized in a way, that includes an entire double page in one image. We aim to create ground truth data for page spanning article separation and explore approaches to separate articles automatically.

	\subsubsection{Enabling Historical Research}
    Ideally, our work will enable the processing of millions of pages of historical data, making vast resources easily
    available to future researchers who can then build upon the transcribed source material, for example, with machine
    translation and \ac{nlp}
    pipelines. Countless research questions concerning economic, societal, political and scientific development can be
    addressed with such data. For a more detailed description of the relevance of such data for historical research, see
    Supplementary Section~\ref{sec:hist_motivation}
    . We hope this dataset will help to improve our understanding of the past.

\end{document}

%% file: math_commands.tex
\usepackage{amsmath,amsfonts,bm}

\def\eqref#1{equation~\ref{#1}}

\def\1{\bm{1}}

\DeclareMathAlphabet{\mathsfit}{\encodingdefault}{\sfdefault}{m}{sl}
\SetMathAlphabet{\mathsfit}{bold}{\encodingdefault}{\sfdefault}{bx}{n}



%% file: acronyms.tex
\DeclareAcronym{cnn}{
    short = CNN,
    long = Convolutional Neural Network,
}
\DeclareAcronym{ocr}{
    short = OCR,
    long = Optical Character Recognition
}
\DeclareAcronym{rnn}{
    short = RNN,
    long = Recurrent Neural Network
}
\DeclareAcronym{lstm}{
    short = LSTM,
    long = Long Short-Term Memory
}
\DeclareAcronym{gpu}{
    short = GPU,
    long = graphics processing unit
}
\DeclareAcronym{iou}{
    short = IoU,
    long = Intersection over Union
}
\DeclareAcronym{nlp}{
    short = NLP,
    long = Natural Language Processing
}
\DeclareAcronym{id}{
    short = iD,
    long = in Distribution
}
\DeclareAcronym{ood}{
    short = OoD,
    long = Out of Distribution
}

%% file: figures/papers_per_year.tex
\begin{tikzpicture}

\definecolor{darkgray176}{RGB}{176,176,176}
\definecolor{steelblue31119180}{RGB}{31,119,180}

\begin{axis}[
/pgf/number format/.cd,
    1000 sep={},
tick align=outside,
tick pos=left,
x grid style={darkgray176},
xlabel={year},
xmin=1656.2, xmax=1981.8,
xtick style={color=black},
y grid style={darkgray176},
ylabel={newspapers},
ymin=-2808.65, ymax=58981.65,
ytick style={color=black},
width=6cm,
height=6cm
]
\addplot [semithick, steelblue31119180]
table {%
1671 2
1672 54
1673 29
1674 0
1675 0
1676 0
1677 0
1678 0
1679 0
1680 0
1681 0
1682 0
1683 0
1684 0
1685 0
1686 0
1687 0
1688 0
1689 0
1690 0
1691 0
1692 0
1693 0
1694 0
1695 0
1696 0
1697 0
1698 0
1699 0
1700 0
1701 0
1702 0
1703 0
1704 0
1705 0
1706 0
1707 0
1708 0
1709 0
1710 0
1711 0
1713 0
1714 0
1715 0
1716 0
1717 0
1718 0
1719 6
1720 0
1721 0
1722 0
1723 0
1724 0
1725 0
1726 0
1727 0
1728 0
1729 0
1730 0
1731 0
1732 0
1733 0
1734 0
1735 0
1736 0
1737 0
1738 0
1739 0
1740 0
1741 0
1742 0
1744 103
1745 0
1747 2
1748 156
1749 3
1750 57
1751 152
1752 105
1753 105
1755 51
1756 103
1757 153
1758 156
1759 99
1760 153
1761 154
1762 100
1763 261
1764 403
1765 660
1766 566
1767 429
1768 532
1769 379
1770 533
1771 523
1772 402
1773 628
1774 744
1775 658
1776 414
1777 594
1778 386
1779 311
1780 743
1781 486
1782 677
1783 673
1784 819
1785 1075
1786 1090
1787 1268
1788 995
1789 1014
1790 749
1791 965
1792 977
1793 1294
1794 1361
1795 1409
1796 1946
1797 1402
1798 1197
1799 1328
1800 1458
1801 1251
1802 1640
1803 1690
1804 1653
1805 1540
1806 1261
1807 1365
1808 1554
1809 2069
1810 2245
1811 2712
1812 2995
1813 2693
1814 3478
1815 3807
1816 3695
1817 3893
1818 3965
1819 3638
1820 3851
1821 4203
1822 4588
1823 4146
1824 5156
1825 5089
1826 5341
1827 5232
1828 5462
1829 5641
1830 6375
1831 6204
1832 6037
1833 6901
1834 5756
1835 6377
1836 6593
1837 7474
1838 7960
1839 8540
1840 7951
1841 8980
1842 9317
1843 9709
1844 9896
1845 10831
1846 10483
1847 10733
1848 12227
1849 13052
1850 12253
1851 13330
1852 12912
1853 12765
1854 12966
1855 13926
1856 13738
1857 13583
1858 13940
1859 14331
1860 14963
1861 15168
1862 15199
1863 15162
1864 15419
1865 15685
1866 14976
1867 14058
1868 14381
1869 14767
1870 15484
1871 14904
1872 16329
1873 17031
1874 17868
1875 19468
1876 19574
1877 18536
1878 20451
1879 22240
1880 23599
1881 24115
1882 25970
1883 26037
1884 27142
1885 28125
1886 30070
1887 29418
1888 31558
1889 31703
1890 34971
1891 35694
1892 36039
1893 37523
1894 38470
1895 38403
1896 40791
1897 41989
1898 43624
1899 42826
1900 44432
1901 44567
1902 45163
1903 47173
1904 47475
1905 50238
1906 49528
1907 50545
1908 52177
1909 52275
1910 50414
1911 51559
1912 51987
1913 52973
1914 56173
1915 55451
1916 54177
1917 53159
1918 51952
1919 49512
1920 48492
1921 48371
1922 44638
1923 40591
1924 44922
1925 46265
1926 47860
1927 48968
1928 48903
1929 48960
1930 46942
1931 48681
1932 48765
1933 44912
1934 43049
1935 40348
1936 38632
1937 38665
1938 37591
1939 34538
1940 33153
1941 27147
1942 26798
1943 24344
1944 16542
1945 2952
1946 893
1947 1066
1948 1298
1949 1939
1950 3579
1951 2736
1952 2796
1953 384
1954 391
1955 399
1956 405
1957 397
1958 403
1959 395
1960 390
1961 383
1962 394
1963 389
1964 378
1965 389
1966 399
1967 380
};
\end{axis}

\end{tikzpicture}

%% file: figures/pipeline/pipeline.tex
\begin{tikzpicture}[module/.style={draw, very thick, rounded corners, minimum width=15ex}, embmodule/.style={module, fill=red!20}, mhamodule/.style={module, fill=orange!20}, lnmodule/.style={
    module, fill=yellow!20}, ffnmodule/.style={module, fill=cyan!20}, , arrow/.style={-stealth, thick, rounded corners},
	line/.style={thick, rounded corners} ]    


    \node[mhamodule, align=center](unet_2) {U-Net\\Inference};
    \node[above=0.2 of unet_2] (baseline) {Baseline};
    \node[right=0.6 of unet_2, module, align=center] (base_labels) {Baseline\\Ascender\\Descendender\\Endpoints};
    \node[below=0 of base_labels, align=center] (regions) {Regions};
    \node[fit={(base_labels)(regions)}, draw, ultra thick, rounded corners, inner sep=6] (base_results) {};
    \node[below=0.2 of unet_2, mhamodule, align=center](peroprocess) {Baseline\\ Postprocessing};

    \draw[arrow] (unet_2) -- (base_labels);
    \draw[arrow] (base_results.west|-peroprocess.east) -- (peroprocess.east);

	\node[fit={(unet_2)(peroprocess)(base_results)}, draw, ultra thick, rounded corners, inner sep=5] (basline_block) {};
	
	\node[above=1.5 of basline_block, embmodule, align=center](unet) {U-Net\\Inference};
	\node[above=0.2 of unet, embmodule, align=center](polprocess) {Layout\\Postprocessing};
	\node[above=0.2 of polprocess] (segmentation) {Segmentation};
	\node[fit={(unet)(polprocess)(segmentation)}, draw, ultra thick, rounded corners] (segmentation_block) {};
	
	\node[right =2.3of segmentation_block] (layout) {\includegraphics[width=.4\textwidth]{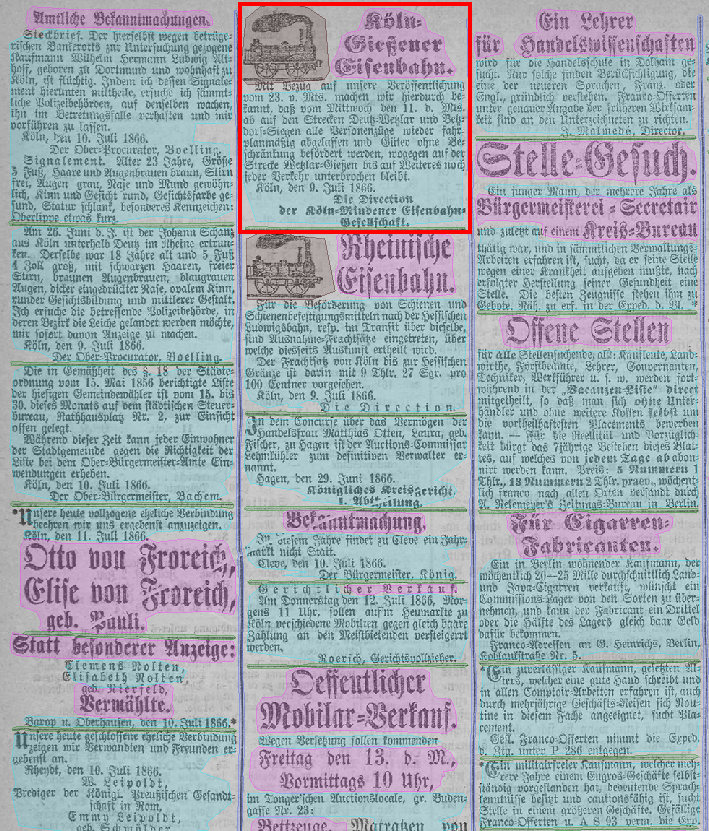}};
	\node[above=0.1 of layout] {Layout Inference};
	\node[left =2.3of segmentation_block] (source) {\includegraphics[width=.4\textwidth]{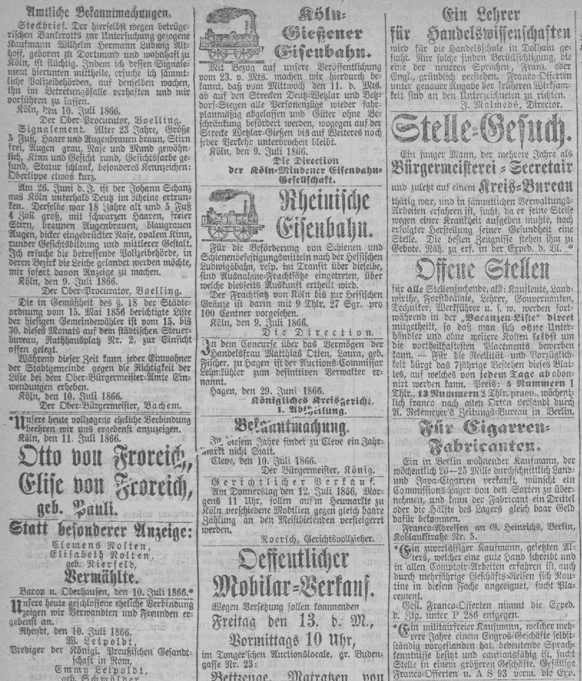}};
	\node[above=0.1 of source] {Input image};
	
	\draw[arrow] (source.east|-unet.west) -- (unet.west);
	\draw[arrow] (unet) -- (polprocess);
	\draw[arrow] (polprocess.east) -- (layout.west|-polprocess.east);
	
    \draw[arrow] (source.south) -- (unet_2.west-|source.south);
	\draw[arrow] (unet_2.west-|source.south) -- (unet_2.west);
	\draw[line] (layout.south) -- (regions.east-|layout.south);
	\draw[arrow] (regions.east-|layout.south) -- (regions.east);
	
	\node[below =0.5 of peroprocess] (pero_check) {};
	\draw[line] (peroprocess.south) -- (pero_check.center);
	
	\node[below=0.7 of basline_block, ffnmodule, align=center](ocr_model) {LSTM\\Inference};
	\node[fit={(unet)(polprocess)(segmentation)}, draw, ultra thick, rounded corners] (segmentation_block) {};

	\node[below =1.5 of peroprocess] (ocr_result_check) {};
	\node[rectangle,draw, left =5.5 of ocr_result_check, align=center] (ocr_result) {Köln-\\
	Gießener \\
	Ei{\fontfamily{jkpvos}\selectfont{}s}enbahn.\\

	Mit Bezug auf un{\fontfamily{jkpvos}\selectfont{}s}ere Veröffentlichung \\
	vom 23. v. Mts. machen wir hierdurch be- \\
	kannt, daß vom Mittwoch den 11. d. Mts. \\
	ab auf den Strecken Deutz-Wetzlar und Betz- \\
	dorf-Siegen alle Per{\fontfamily{jkpvos}\selectfont{}s}onenzüge wieder fahr- \\
	vlanmäßig abgela{\fontfamily{jkpvos}\selectfont{}s}{\fontfamily{jkpvos}\selectfont{}s}en und Güter ohne Be- \\
	{\fontfamily{jkpvos}\selectfont{}s}chränkung befördert werden, wogegen auf der \\
	Strecke Wetzlar-Gießen bis auf Weiteres noch \\
	jeder Verkehr unterbrochen bleibt. \\
	Köln, den 9. Juli 1866. \\
	Die Direction
	der Köln-Mindener Ei{\fontfamily{jkpvos}\selectfont{}s}enbahn- \\
	Ge{\fontfamily{jkpvos}\selectfont{}s}ell{\fontfamily{jkpvos}\selectfont{}s}chaft.};
	\node[above=0.1 of ocr_result] {Optical Character Recognition};
	\node[right =8.5 of ocr_result_check] (lines) {\includegraphics[width=.4\textwidth]{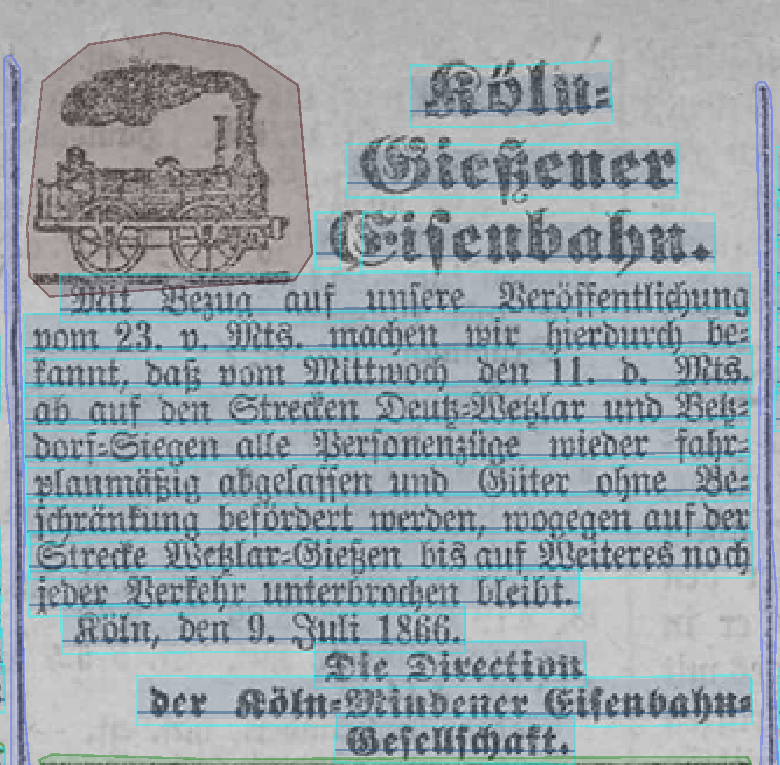}};
	\node[above=0.1 of lines] {Text-Line Inference};
	
	\draw[arrow] (pero_check.center) -- (lines.west|-pero_check.center);
	
	\draw[line] (unet_2.west-|source.south) -- (source.south|-ocr_model.west);
	\draw[arrow] (source.south|-ocr_model.west) -- (ocr_model.west);
	\draw[arrow] (lines.west|-ocr_model.east) -- (ocr_model.east);
	
	\node[below =0.3 of ocr_model] (ocr_model_check) {};
	\draw[line] (ocr_model.south) -- (ocr_model_check.center);
	\draw[arrow] (ocr_model_check.center) -- (ocr_model_check.center-|ocr_result.east);

\end{tikzpicture}

%% file: figures/legend/legend.tex
\pagestyle{empty}

\def\sep{1.1cm}

\definecolor{paragraph}{HTML}{1ce6ff}
\definecolor{heading}{HTML}{ff34ff}
\definecolor{caption}{HTML}{00ff00}
\definecolor{header}{HTML}{006fa6}
\definecolor{seperatorvertical}{HTML}{3b5dff}
\definecolor{table}{HTML}{ffff00}
\definecolor{image}{HTML}{400000}
\definecolor{inverted_text}{HTML}{ff2f80}
\definecolor{separator}{HTML}{008941}

\begin{tikzpicture}

    \node at (0, 0) {paragraph};
    \draw [draw=paragraph,fill=paragraph] (1.5,-0.25) rectangle ++(.5,.5);

    \node at (0, -1) {heading};
    \draw [draw=heading,fill=heading] (1.5,-1.25) rectangle ++(.5,.5);

    \node at (0, -2) {caption};
    \draw [draw=caption,fill=caption] (1.5,-2.25) rectangle ++(.5,.5);

    \node at (0, -3) {header};
    \draw [draw=header,fill=header] (1.5,-3.25) rectangle ++(.5,.5);

    \node at (0, -4) {seperator\_vertical};
    \draw [draw=seperatorvertical,fill=seperatorvertical] (1.5,-4.25) rectangle ++(.5,.5);

    \node at (0, -5) {table};
    \draw [draw=table,fill=table] (1.5,-5.25) rectangle ++(.5,.5);

    \node at (0, -6) {image};
    \draw [draw=image,fill=image] (1.5,-6.25) rectangle ++(.5,.5);

    \node at (0, -7) {inverted\_text};
    \draw [draw=inverted_text,fill=inverted_text] (1.5,-7.25) rectangle ++(.5,.5);

    \node at (0, -8) {separator\_horizontal};
    \draw [draw=separator,fill=separator] (1.5,-8.25) rectangle ++(.5,.5);

\end{tikzpicture}

%% file: guidelines/guide_text.tex
\subsubsection{Introduction}
These annotation guidelines are an adaptation of the OCR-D rules (
\url{https://ocr-d.de/en/gt-guidelines/trans/transkription.html}
). We outline additional rules, we created to ensure consistency of the \textit{Chronicling Germany} dataset.

\subsubsection{Page types and type area}
The OCR-D guidelines provide for a distinction to be made between page types and the type area during layout
analysis. The type area usually contains the text body, but not elements such as the page number. In the
\textit{Chronicling Germany} data set, these steps are currently not taken into account.

\subsubsection{Regions}
\textbf{Region-types}\label{sec:region_types}
The OCR-D guidelines distinguish between different types of regions, such as text, image and separator regions. In
the Bonn Newspaper dataset, the regions are generally recorded in accordance with OCR-D page region level 1 (
\url{https://ocr-d.de/de/gt-guidelines/trans/ly_level_1_5.html}
). However, tables are also recorded as a separate region and no distinction is made between images and drawings;
instead, all images, photos, illustrations and drawings are grouped together under the GraphicRegion. The entire
contiguous region is always marked as a block. For text regions, this applies to contiguous blocks of the same class.

\begin{itemize}
    \item \texttt{TextRegion:} All texts that are not tables. Table headings are not marked as a text region.
    \item \texttt{TableRegion:}
    All parts of the page that contain tabular information. These are often, but not always, clearly recognizable as
    tables by small separators. Text that is only separated by separators does not count as a table, but a structure
    must be recognizable that assigns certain meanings to rows and columns. Table headings are included with
    corresponding tables.
    \item \texttt{SeparatorRegion:}
    All dividing lines are marked as SeparatorRegion. This also includes decorative elements that, like other
    separator lines, separate areas from each other and are not purely cosmetic in nature. The separators are
    divided into vertical and horizontal separators and marked with “separator\_vertical” and “separator\_
    horizontal”.
    \item \texttt{GraphicRegion:} All graphics, images, photos, illustrations, and drawings.
\end{itemize}

\subsubsection{TextRegion subtypes}
\label{sec:textregion}
TextRegions are divided into different subtypes. The subdivision corresponds to the OCR-D guideline for text regions
(\url{https://ocr-d.de/de/gt-guidelines/trans/lytextregion.html#textregionen__textregion_}
). However, drop capitals are treated differently from OCR-D. These are counted as part of the paragraph instead of
being marked as a separate text region so that models trained on this data will include them in the correct position
in their text output. In addition, headlines (caption) and inverted text (inverted-text) are also recorded in the
\textit{Chronicling Germany}
data set. Instead of annotating advertisements separately, the classes created for other newspaper pages are applied
to the advertisements as far as possible. Because headlines should be visually identified, this leads to a large
number of text in the advertisements marked as headlines, which contradicts a semantic definition of a headline.
Therefore, it makes sense to treat these pages separately in practice and not differentiate between headings and
other text. The following elements from the OCR-D guidelines are not represented in the \textit{Chronicling Germany}
dataset due to lack of occurrence:\\
page-number, marginalia, footnote, signature-mark, catch-word, floating, TOC-entry

We discuss the definition for the text subclasses below:
\begin{itemize}
    \item \textbf{paragraph:}
    Standard text type that includes paragraphs. These are usually kept compact to accommodate as much text as
    possible in the available space. If a text region cannot be assigned to any other type,
    it falls under the paragraph label.
    \item \textbf{heading:}
    Headings that can be clearly distinguished visually from the rest of the text. This is achieved by using a
    significantly larger or bold font and centered text, which is clearly different from the block layout of
    paragraphs. A heading is located above a paragraph and is sometimes separated from the previous text by a
    separator. A thin separator between the heading and the text can occur. However, if there is too much space
    between them or a thick separator, the two texts no longer count as belonging together in the sense of heading
    and paragraph. If a text is not superordinate to a paragraph, it cannot be a heading.
    \item \textbf{header:}
    Page or column titles that appear prominently above the entire page. These are centered at the top of the page
    and can appear in different font sizes.
    \item \textbf{caption:}
    Title lines that are located to the right and left of a page heading or text heading. They often contain
    information such as the date.
    \item \textbf{inverted-text:}
    Text that is printed white on black. This is often part of decorative elements but is not marked as a graphic
    element.
\end{itemize}

\subsubsection{\ac{ocr}}
A prerequisite for text recognition is baseline or text-line recognition. Both the baseline itself and a polygon
around the text line are annotated. These are generated automatically and only corrected if the baseline connects
non-contiguous text passages. Lines that have been divided into two baselines are not corrected. Tables and inverted
text are not given baselines.

The text is corrected according to its optical appearance. What is written on the page is transcribed, even if there
are errors in the print or scan. Completely illegible passages are not transcribed, passage that are largly illegible are transcribed but marked with the "unclear" tag of Transkibus. These passage are not used in training.

Zodiac signs and geometric signs at the beginning of some paragraphs – a peculiarity of the Kölnische Zeitung – are not transcribed.

The transcription is carried out according to level 2 of the OCR-D guidelines (
\url{https://ocr-d.de/en/gt-guidelines/trans/level_2_2.html}
). This includes the transcription of special characters such as the 'long s' (U+017F) or long hyphens (U+2014, em
dash). Consistency with the rest of the data is important here. As these were generated automatically, it is best to
look for another example and adopt that version if the special characters are unclear.

Unlike in the OCR-D guideline, fractions are not transcribed with special characters. Instead, the fraction is
represented with a slash:\\ $1 \frac{3}{4}$ = 1 3/4.\\
In this case, it is important to separate the whole number from the fraction with a space. The same applies to times
with an underscore. Example for clock times: $11_{45} = $11\_45 or $11._{45} = $11.\_
45. (For both, use non-breaking spaces in future (U+202F))

The case of a number that is followed by a unit (e.g. 100M) is not dissolved in the OCR-D guidelines. We always add a space between the number and the unit (e.g. 100M becomes: 100 M).

The units of the Lübische Mark currency are transcribed as follows: the Lübische Mark as Ml., the Lübische Pfennig as dl. and the Lübische Schilling as ßl., where 1 Ml. = 16 ßl. = 192 dl. (\url{https://www.hagen-bobzin.de/hobby/muenzverein_wendisch.html}).

\begin{table}[h]
    \centering
    \begin{tabular}{cc}
       \includegraphics[]{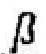}  & bl.\\
        \includegraphics[]{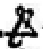} & Ml.\\
    \end{tabular}
    \label{tab:my_label}
\end{table}

Transkribus allows the selection of special characters with a virtual keyboard. However, it must be ensured that the
character used is unique. For example, U+2014 and U+2015 are visually indistinguishable. U+2014 must be used for
long hyphens. If the characters are unclear, the OCR-D guidelines, which include tables for the use of special
characters, can also be consulted:
\begin{itemize}
    \item \url{https://ocr-d.de/en/gt-guidelines/trans/trLigaturen2.html}
    \item \url{https://ocr-d.de/en/gt-guidelines/trans/trFremdsprache.html}
    \item \url{https://ocr-d.de/en/gt-guidelines/trans/ocr_d_koordinationsgremium_codierung.html}
    \item \url{https://ocr-d.de/en/gt-guidelines/trans/trBeispiele.html}
    \item \url{https://ocr-d.de/en/gt-guidelines/trans/tr_level_1_3.html}
    \item \url{https://ocr-d.de/en/gt-guidelines/trans/trAnfZeichen.html}
    \item \url{https://ocr-d.de/en/gt-guidelines/trans/trGedankenstrich.html}
\end{itemize}

%% file: figures/legend/legend_caption_heading_paragraph.tex
\pagestyle{empty}

\def\sep{1.1cm}

\definecolor{paragraph}{HTML}{1ce6ff}
\definecolor{heading}{HTML}{ff34ff}
\definecolor{caption}{HTML}{00ff00}
\definecolor{header}{HTML}{006fa6}
\definecolor{seperatorvertical}{HTML}{3b5dff}
\definecolor{table}{HTML}{ffff00}
\definecolor{image}{HTML}{400000}
\definecolor{inverted_text}{HTML}{ff2f80}
\definecolor{separator}{HTML}{008941}

\begin{tikzpicture}

    \node at (0, 0) {paragraph};
    \draw [draw=paragraph,fill=paragraph] (1.5,-0.25) rectangle ++(.5,.5);

    \node at (0, -1) {heading};
    \draw [draw=heading,fill=heading] (1.5,-1.25) rectangle ++(.5,.5);

    \node at (0, -2) {caption};
    \draw [draw=caption,fill=caption] (1.5,-2.25) rectangle ++(.5,.5);

\end{tikzpicture}

%% file: figures/legend/legend_heading_header.tex
\pagestyle{empty}

\def\sep{1.1cm}

\definecolor{paragraph}{HTML}{1ce6ff}
\definecolor{heading}{HTML}{ff34ff}
\definecolor{caption}{HTML}{00ff00}
\definecolor{header}{HTML}{006fa6}
\definecolor{seperatorvertical}{HTML}{3b5dff}
\definecolor{table}{HTML}{ffff00}
\definecolor{image}{HTML}{400000}
\definecolor{inverted_text}{HTML}{ff2f80}
\definecolor{separator}{HTML}{008941}

\begin{tikzpicture}

    \node at (0, 0) {header};
    \draw [draw=header,fill=header] (1.5,-0.25) rectangle ++(.5,.5);

    \node at (0, -1) {heading};
    \draw [draw=heading,fill=heading] (1.5,-1.25) rectangle ++(.5,.5);

\end{tikzpicture}